\documentclass[%
aps,
prapplied,
amsmath,
amssymb,
superscriptaddress,
reprint,
floatfix, 
longbibliography,
]{revtex4-2}
\usepackage{graphicx}
\usepackage{siunitx}
\usepackage[dvipsnames]{xcolor}
\usepackage{mathtools}
\usepackage{textcomp}
\usepackage[overload]{empheq}
\usepackage{mleftright}
\usepackage{xparse}
\usepackage{cancel}
\usepackage{amsmath}
\usepackage{accents}
\usepackage{relsize}
\usepackage{lmodern}
\usepackage{mathrsfs}
\usepackage{textcomp}
\usepackage[hidelinks,colorlinks=true,linkcolor=blue,citecolor=blue,urlcolor=blue]{hyperref}

\NewDocumentCommand{\evalat}{sO{\big}mm}{%
  \IfBooleanTF{#1}
   {\mleft. #3 \mright|_{#4}}
   {#3#2|_{#4}}%
}
\newcommand{\FReac}[0]{$F\delta_0^\mathrm{reac}$~}
\newcommand{\FDiss}[0]{$F\delta_0^\mathrm{diss}$~}
\newcommand{\nbar}[0]{$\overline{n}$~}
\newcommand{\Teff}[0]{$T_\mathrm{eff}$~}
\newcommand{\ubar}[1]{\underaccent{\bar}{#1}}

\begin{document}


\title{Thin-film quartz for high-coherence piezoelectric phononic crystal resonators} 

\author{A. L. Emser}
\email{alec.emser@colorado.edu}
\author{C. Metzger}
\author{B. C. Rose}

\affiliation{JILA, National Institute of Standards and Technology and the University of Colorado, Boulder, Colorado 80309, USA}
\affiliation{Department of Physics, University of Colorado, Boulder, Colorado 80309, USA}

\author{K. W. Lehnert}
\affiliation{JILA, National Institute of Standards and Technology and the University of Colorado, Boulder, Colorado 80309, USA}
\affiliation{Department of Physics, University of Colorado, Boulder, Colorado 80309, USA}
\affiliation{Department of Physics, Yale University, New Haven, Connecticut 06511, USA}

\date{\today}

\begin{abstract}

Piezoelectric phononic crystal resonators (PCRs) are a promising platform for acoustic quantum processing, yet their performance is currently limited by coupling to an ensemble of saturable two-level system (TLS) defects within the resonator material. Motivated by its excellent bulk mechanical properties and high crystallinity, we address this by fabricating PCRs from a new substrate: thin-film quartz. At single-phonon powers and millikelvin temperatures -- requisite conditions for quantum phononic processing -- we demonstrate large internal mechanical quality factors, $Q_i > 160,000$. This represents an order of magnitude improvement in single-phonon lifetimes for piezoelectric PCR. We characterize the loss channels in these devices and find that, although improved, the low-power response is still limited by coupling to a TLS bath and that a significant portion of the TLSs are associated with the aluminum coupling electrodes. To explore the high-power response we perform ringdown measurements and demonstrate high-power quality factor-frequency products $Q_i \cdot f = 1.4 \times 10^{16}$ Hz. 
\end{abstract}

\maketitle 

\section{Introduction}

By integrating long-lived mechanical resonators with highly nonlinear superconducting qubits, circuit quantum acoustodynamics (cQAD) has demonstrated sophisticated quantum control of phonons \cite{Chu2018, Bild2023, Sletten2019,Arrangoiz2018, Arrangoiz2019, Lupke2022, Moores2018, Lee2023, Malik2023, qiao2023}. With continued development, these hybrid superconducting-phononic systems may facilitate efficient quantum processing, storage, and networking \cite{Pechal2018, Hann2019, arrangoiz2019IEEE, Chamberland2022}. At present, however, the utility of many cQAD platforms is limited by qubit lifetimes which are orders of magnitude shorter than those of state-of-the-art qubits which are not integrated with acoustic resonators. This undesirable decay in hybrid systems results in part from the spontaneous emission of the qubit into mechanical degrees of freedom \cite{Jain2023, Kitzman2023}. To minimize this acoustic emission it is paramount that not only must the qubit couple to modes of the mechanical resonator which are long-lived, but also that these modes should be sufficiently isolated from others to prevent inadvertent coupling between a qubit and spurious mechanical modes. 

In the search for a mechanical resonator which satisfies these criteria, the phononic crystal resonator (PCR) has emerged as a strong candidate. By placing a defect cell between arrays of phononic shields engineered to support an acoustic bandgap, PCRs have demonstrated long-mechanical lifetimes \cite{MacCabe2020} and a suppressed acoustic density of states \cite{Wollack2021, Chen2023}. Furthermore, fabricating PCRs on piezoelectric substrates enables direct coupling between the mechanical defect mode and superconducting qubits. However, such devices fabricated on piezoelectric thin-films typically suffer from large mechanical loss at single-phonon power and millikelvin temperature; previous work has demonstrated single-phonon coherence times corresponding to internal quality factors $Q_i\approx11,000-17,000$ \cite{Wollack2022,Lee2023}. This decay is suspected to be dominated by coupling to a bath of microscopic two-level systems (TLS) \cite{MacCabe2020}, the magnitude of which is heavily influenced by choices of material and fabrication processes \cite{Crowley2023,Gruenke2023, Mittal2024}.

The high crystallinity, excellent bulk mechanical properties, and extensive technical characterization of quartz position it as an attractive alternative to commonly used thin-film piezoelectrics \cite{Goryachev2013, Landau1937, Emser2022, Kharel2018}.  Although the growth of high quality bulk monocrystalline quartz is now a mature and industrial-scale process \cite{Spezia1909, Hale1948, Brice1985}, only recently have efforts focused on producing thin-film quartz \cite{stratton2004, Kubena2006, Carretero2013,Zhang2019, Jolly2020, Lutjes2021,antoja2021} and its applications to microelectromechanical systems remain largely unexplored. Motivated to create a platform that hosts spectrally isolated and low-loss mechanical modes at low temperature and power, thin-film quartz thus emerges as a promising substrate for quantum acoustics.
 
In this work, we fabricate phononic crystal resonators from a 1 \textmu m thin film of quartz on silicon. We demonstrate high internal mechanical quality factors ($Q_i>160,000$) at mK temperatures and single-phonon intracavity occupancies. We characterize the loss mechanisms in these devices and demonstrate that the dominant source of low-temperature and low-power dissipation arises from coupling to a TLS bath, the magnitude of which is correlated with mechanical participation in the aluminum that defines the resonator electrodes. We disentangle the intrinsic TLS loss tangents of the quartz and aluminum  in these devices and find $\delta_{qz}$ = $4.5\pm1.6 \times 10^{-6}$ and $\delta_{Al}$ = $4.9\pm1.0 \times 10^{-4}$, respectively. Notably, we are unable to resolve viscoelastic damping \cite{Wollack2021} or any other form of power- or temperature-independent dissipation at low drive powers and temperatures, and the measured quality factors represent an order of magnitude improvement for PCRs in this regime \cite{Wollack2021, Wollack2022, Lee2023}. We additionally use ringdown measurements to characterize these thin-film quartz PCRs at high powers for which we demonstrate large quality factor-frequency products $Q_i \cdot f = 1.4 \times 10^{16}$ Hz.

\section{phononic crystal resonators}
Acoustic bandgap engineering relies on periodically patterning an elastic host material to create a frequency band inside of which phononic excitations are suppressed. PCRs leverage this technique by introducing a defect site inside of the periodic structure such that a mechanical resonance frequency of the defect site lies within the bandgap. This mode is thereby locally confined to the defect site and its decay into the surrounding material is heavily suppressed by the phononic shielding. Our devices begin with a 1 \textmu m thin layer of quartz bonded to 500 \textmu m of silicon (Appendix \ref{sec:apdx_fabrication}). Each PCR consists of two arrays of crenellated phononic shields which sandwich a defect site to form a thin-film quartz bridge. This bridge is released from the silicon below to realize a freestanding structure. To excite the defect mode an aluminum coupling capacitor is patterned on the surface of the defect site. The leads on each side of the coupling capacitor traverse the phononic shielding to reach the signal and ground of a coplanar waveguide transmission line (CPW). By shunting multiple resonators to one CPW and offsetting their resonant frequencies, we facilitate the study of many mechanical resonances with one device \cite{Wollack2021}.

\begin{figure*}
\centering
\includegraphics[width=\textwidth]{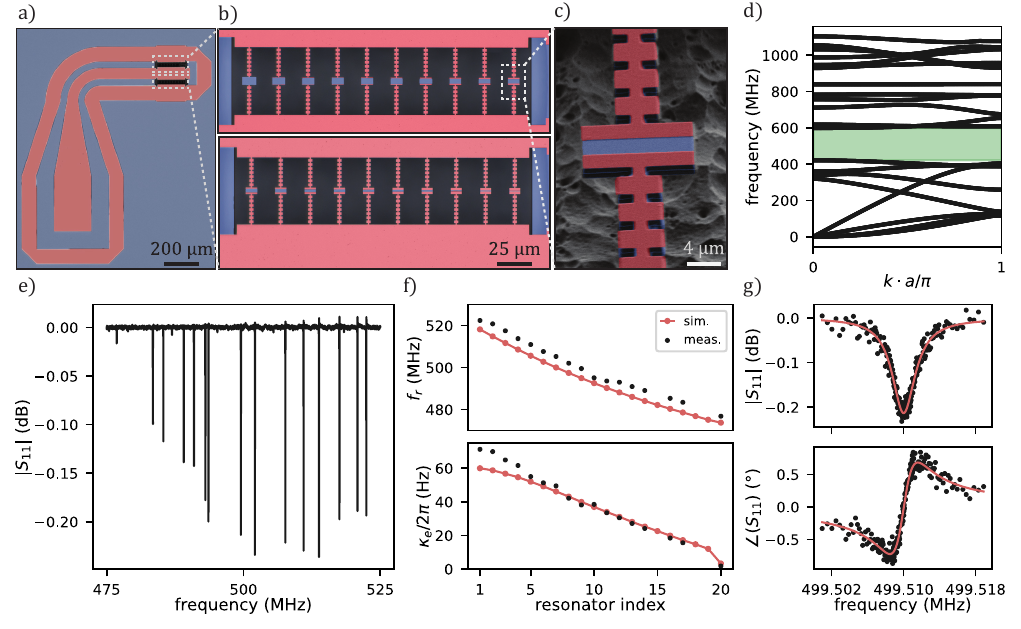}
\caption{Thin-film quartz phononic crystal resonators. (a-c) False-colored scanning-electron micrographs (SEM) of the device comprising twenty suspended PCR shunting one open end of a microwave CPW transmission line probed in reflection. Each device consists of a 50 nm layer of aluminum (red) deposited on 1 \textmu m of thin-film quartz (blue) bonded to a silicon (black) wafer. The PCR are defined by etching the quartz into a series of one-dimensional beams and releasing them from the silicon substrate by means of an XeF\textsubscript{2} etch. (b) SEM image of the twenty PCR showing how both the amount of aluminum on each defect site and the size of the defect cell is swept over the two rows of devices. (c) Zoom-in on the defect site of one suspended PCR: two aluminum electrodes running atop the phononic shields are used to excite a mechanical mode of the defect site. (d) Simulated phononic bandgap for the arrayed phononic shields with unit cell length $a$ as we sweep the wavevector $k$ over the one-dimensional Brillouin zone. A complete bandgap (green) is expected in the range 421-598 MHz. (e) Magnitude of the reflection coefficient $S_{11}$ as a function of probe frequency showing 17 strongly coupled resonances centered around $500$~MHz.  (f) The defect mode resonance frequencies (top) and external couplings (bottom) as predicted by FEM simulations (red) and as measured (black markers). (g) Magnitude (top) and phase (bottom) of $S_{11}$ around one particular resonance, probed at low microwave power resulting in $\overline{n}=1.4 \pm 0.07$ and $Q_i = 161,000 \pm 3,600$. Experimental data is shown with black markers and the solid red line shows the fit. }
\label{fig:fig1}
\end{figure*}

In Fig.~\ref{fig:fig1}, we show the primary device studied in this work, including (a) the CPW, (b) arrays of mechanical resonators, and (c) a zoomed-in view of one PCR. The bridge for each PCR consists of seven phononic shields on each side of the defect site from which we expect a phononic bandgap spanning 421-598 MHz, as shown in Fig. \ref{fig:fig1}(d). The device consists of 20 PCR connected to the CPW in parallel. Over the array, two design parameters are swept: the area of the defect cell and the width of the gap between the coupling electrodes. Sweeping the area of the defect cell tunes the frequency of the defect mode while adjusting the electrode gap will vary the resonator external coupling rate $\kappa_e$ as well as the amount of aluminum in which mechanical energy will participate. We design the 20 defect modes to be spaced evenly between 480 and 520 MHz, with the lowest frequency resonators having the smallest external coupling (largest electrode gap) and the highest frequency resonators having the largest external coupling (smallest electrode gap). The device is mounted to the base plate of a dilution refrigerator with lowest operating temperature 8 mK. 

We locate the defect mode resonance frequencies $f_r$ by measuring the reflection of a continuous microwave tone of frequency $f$ over a broad range $475<f<525$ MHz.  As shown in Fig. \ref{fig:fig1}(e), the magnitude of the reflection coefficient, $S_{11}(f)$, features 17 dips, each corresponding to a mode of one of the defect resonators. Although we expect 20 resonances, it is likely that complications in fabrication resulted in several unresponsive resonators. With the resonant frequencies identified, we proceed by measuring further traces over narrower frequency bands to characterize each resonance individually. The reflection coefficient $S_{11}(f)$ around a resonance takes the form of a circle in the complex plane which we fit using the diameter correction method (DCM),
\begin{equation}\label{eq:lorentzianFit}
    S_{11}(f) = 1-e^{i\phi}\frac{2Q/|\hat{Q_e}|}{1+2iQ(f-f_r)/f_r},
\end{equation}
where $\hat{Q}_e=Q_e e^{-i\phi}$ is the complex external quality factor with phase $\phi$, and the total quality factor is given by $Q = (Q_i^{-1}+Q_e^{-1})^{-1}$ for which $Q_i$ is the internal quality factor and the DCM-adjusted external quality factor is $Q_e = |\hat{Q_e}|/\mathrm{cos}(\phi)$ \cite{Khalil2012}. Fitting the complex-valued $\hat{Q}_e$ accounts for a potential rotation $\phi$ of the resonance curve in the complex plane induced by impedance mismatches which may otherwise result in overestimation of $Q_i$ \cite{Khalil2012}. The internal (external) coupling rates are subsequently calculated as $\kappa_{i(e)}=2\pi f_r/Q_{i(e)}$. We fit the reflection response of each resonance and compare in Fig. \ref{fig:fig1}(f) the measured values of $f_r$ and $\kappa_e$ with the ones obtained from finite element method (FEM) simulations \cite{COMSOL}. We find that the simulations demonstrate strong predictive ability; the measured values of $f_r$ are all found within $1\%$ of the simulated values and the external coupling rates are predicted within $\pm 11\%$.

We are particularly interested in the response of the PCRs with small driving powers, a requisite condition for operation in a hybrid cQAD system. On resonance, the mean intracavity phonon occupancy is given by
\begin{equation}\label{eq:intracavPhonons}
    \overline{n} = 4\frac{\kappa_e}{\kappa^2}\frac{P_s}{hf_r},
\end{equation}
where $P_s$ is the microwave power at the sample and the total linewidth $\kappa$ is defined as $\kappa = \kappa_i + \kappa_e$. In Fig. \ref{fig:fig1}(g), we show the response of the $f_r = 499.5$ MHz resonance at a drive power corresponding to $\overline{n}=1.4 \pm 0.07$. From the fit we find that $Q_i = 161,000 \pm 3,600$, which represents an order of magnitude improvement in the single-phonon coherence for a piezoelectric PCR at millikelvin temperatures \cite{Wollack2021,Wollack2022, Lee2023}. The remainder of this work explores the dominant dissipation mechanisms for these devices. 

\section{Dissipative TLS sources}\label{sec:characLoss}
TLS defects are suspected to dominate sources of dissipation and decoherence in a myriad of superconducting quantum systems \cite{Martinis2005, Gao2008, MacCabe2020, Spiecker2023, Muller2019} including piezoelectric PCRs \cite{Wollack2021, Wollack2022}. Accordingly, significant effort in recent years has focused on mitigating the deleterious effects induced by coupling to these defects \cite{Gruenke2023, Crowley2023, Chen2023}. TLSs are typically described within the framework of the standard tunneling model \cite{Phillips1987} which was developed to describe thermal transport in amorphous solids at low temperatures \cite{Phillips1972, Anderson1972, Jackle1972}. The model assumes that a TLS can occupy one of two energetically similar configurations -- modeled as minima in an asymmetric double-well potential -- and may couple to external electric or strain fields. The interaction between a localized TLS and the strain field of a mechanical resonator is usually modeled through linear response theory, which describes the resonator response in terms of a complex mechanical susceptibility $\chi$. The coupling induces both dissipative (imaginary) and reactive (real) contributions to $\chi$ which are measured by monitoring the resonator loss and frequency, respectively. Although at low temperatures the TLS coupling is typically dominated by \textit{resonant} interactions with TLSs which transition near the resonator frequency, at higher temperatures TLSs far off resonance may also affect $\chi$ through a process known as \textit{relaxation} damping. Understanding the contributions of both forms of TLS coupling is critical for realizing a full characterization of the device loss. 

Resonant TLS coupling arises from dispersive interactions between the mechanical resonator and TLSs near $f_r$. We model the resonant-TLS dissipation as
\begin{equation}\label{eq:resTLS_Q}
    Q_\mathrm{res}^{-1} = {F\delta_0^\mathrm{diss}}\frac{\mathrm{tanh}(\frac{hf}{2k_\mathrm{B}T})}{\sqrt{1+(\frac{\overline{n}}{n_c})^\beta\mathrm{tanh}(\frac{hf}{2k_\mathrm{B}T})}},
\end{equation}
where $T$ is the device temperature, $n_c$ is the critical intracavity phonon number for TLS saturation, $\beta$ is a phenomenological parameter to describe spatially non-uniform TLS saturation \cite{Wang2009}, $\delta_0^\mathrm{diss}$ is the intrinsic TLS loss tangent, and $F$ is the filling fraction of the TLSs in the host material. The term $\mathrm{tanh}(\frac{hf}{2k_\mathrm{B}T})$ appearing in the denominator accounts for the temperature-dependence of the TLS decay time \cite{Crowley2023}. The product \FDiss is commonly used as a figure of merit to characterize a system's TLS response; its inverse provides the quality factor expected from the resonator in the limit of zero temperature and no driving power. Resonant damping is expected to dominate at low temperatures and drive powers for which the TLS bath is predominantly in its ground state.

By contrast, TLS relaxation damping dominates contributions to $\chi$ at temperatures for which the bath is sufficiently thermally populated. As strain and electric fields from the resonator couple to far off-resonant TLSs and perturb their energies, the TLS ensemble will be displaced from thermal equilibrium and an ensuant relaxation process will manifest as decay in the coupled resonator. Because this equilibration process is highly dependent on phonon-phonon interactions, the effective dimensionality $d$ of the surrounding phonon bath will restrict the rate at which relaxation damping occurs. Following previous works \cite{Behunin2016, MacCabe2020,Wollack2021}, we model this relaxation damping as
\begin{equation}\label{eq:relTLS_Q}
    Q_\mathrm{rel}^{-1} = \frac{1}{Q_\mathrm{rel}^{T_0}}\Bigl(\frac{T}{T_0}\Bigr)^d,
\end{equation}
where $T_0$ is an arbitrary temperature at which the relaxation damping introduces loss corresponding to a quality factor $Q_\mathrm{rel}^{T_0}$. For a full derivation of the TLS model, see Appendix \ref{apdx: TLS model}.

\begin{figure}
\centering
\includegraphics[width=\columnwidth]{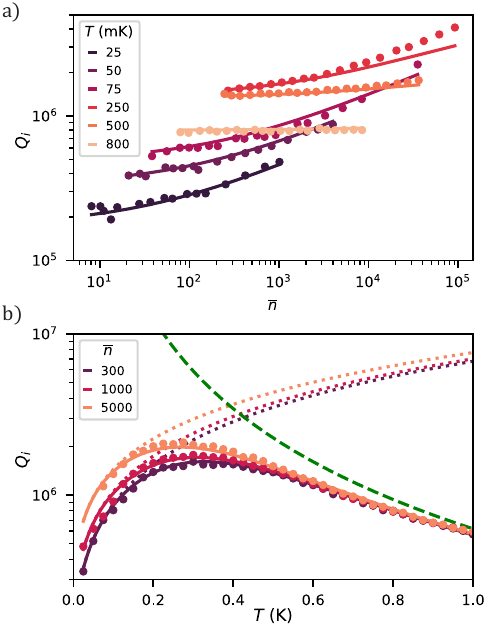}
\caption{Power and temperature dependent loss for the 499.5 MHz resonance. (a) $Q_i$ is plotted against $\overline{n}$ for select device temperatures shown in shades of black-orange.  The two-dimensional fit to the joint model (\ref{eq:jointQModel}) is shown with solid lines. (b) Slices of data corresponding to approximate values of selected $\overline{n}$ are shown with shades of black-orange. Fits to the joint model are shown with solid lines while the resonant-TLS contribution is shown with colored, dotted lines and the power-independent relaxation-TLS contribution is shown with a dashed green line. We emphasize that each fit line is derived from the same set of fitting parameters. Error bars corresponding to one standard deviation of fit uncertainty in $Q_i$ are smaller than the marker size in both subfigures.}
\label{fig:fig2}
\end{figure}

We disentangle these two regimes of dissipative TLS contributions by studying one PCR over a broad range of temperatures and drive powers. The base plate temperature of the refrigerator $T$ is stepped from 25-1000 mK and the power applied to the sample $P_s$ is swept over a 21 dB range. For each combination of temperature and drive power we measure the complex reflection coefficient around the 499.5 MHz resonance and each of these traces is fit to (\ref{eq:lorentzianFit}) to produce a two-dimensional data set $Q_{i}(\overline{n},T)$. To ensure the accurate calculation of $P_s$ and $\overline{n}$, we measure the gain and attenuation of the measurement lines by means of a variable-temperature stage with a calibrated noise source (Appendix \ref{sec:apdx_VTScalib}). We jointly fit the resonator loss to the model
\begin{equation}\label{eq:jointQModel}
    Q_{i}^{-1} = Q_\mathrm{res}^{-1}(\overline{n},T) + Q_\mathrm{rel}^{-1}(T) + Q_\mathrm{bkg}^{-1},
\end{equation}
where the quantity $Q_\mathrm{bkg}^{-1}$ is intended to account for any power and temperature independent losses.

In Fig. \ref{fig:fig2}(a), we plot the measured and fit values $Q_i(\overline{n})$ for selected values of $T$. At low drive powers and temperatures for which the TLS bath thermal occupation is small, the decay of the resonator into near-resonant TLSs dominates the loss. As the drive power is increased, however, resonant coupling gradually saturates the TLS ensemble and improves $Q_i$.

In Fig. \ref{fig:fig2}(b) we highlight the temperature dependence of the TLS bath response by plotting $Q_i(\overline{T})$ for the nearest-available values ($\pm 10\%$) of selected $\overline{n}$. At low $T$, increasing the bath temperature has the same effect as increasing the strength of the drive tone applied at the resonator frequency: near-resonant TLSs become saturated and the resonant damping is suppressed. As the bath temperature is raised further, the resonator field perturbs an increasing number of thermally-occupied TLSs which contribute to relaxation damping. This mechanism competes with the $Q_i$ improvement from resonant TLS saturation and results in the turnover observed at $250$ mK. At sufficiently high temperatures, relaxation damping dominates the resonator dissipation and there is no discernible power dependence of the resonator response because the contributing TLSs are predominantly far detuned from $f_r$.

The fit parameters from this analysis are listed in Table \ref{tab:table1}. From these data we are unable to fit a meaningful value for $Q_\mathrm{bkg}^{-1}$, suggesting that the resonator decay at mK temperatures and low drive powers is entirely captured by the model for coupling to a TLS ensemble. This implies that the quartz PCR performance in the low power regime is not appreciably limited by radiative leakage through the phononic shields (Appendix \ref{sec:apdx_radLeakage}) or by viscoelastic losses resulting from mechanical participation in the aluminum electrodes as reported elsewhere \cite{Wollack2021}.

We note that although $1/F\delta_0^\mathrm{diss}$ is a factor of two lower than the value of $Q_i$ reported for the trace shown in Fig. \ref{fig:fig1}(g), these data were measured on separate cooling cycles of the fridge and we do not expect the properties of the TLS bath coupling to remain the same between thermal cycles \cite{Chen2023}. At intermediate temperatures ($T\approx 250$ mK) and very high drive powers ($\overline{n} > 10,000$) for which the resonant loss contribution is smallest, the measured values of $Q_i$ exceed those predicted by the fit; this is evidenced in Fig. \ref{fig:fig2}(a). We interpret this departure as the effect of readout-power heating of the device: at such high power, the effective device temperature may exceed the mixing chamber plate temperature. At low temperatures ($T<100$ mK) and high drive powers ($\overline{n} \gtrsim 5,000$), the resonant lineshape can no longer be modeled by (\ref{eq:lorentzianFit}) due to a nonlinearity resulting from the TLS bath \cite{Metzger2024:inPrep}. In particular, the heating induced by the probe tone significantly alters the reactive and dissipative loads which the TLS bath imparts to the resonator as the tone is swept across the resonance; we exclude data with such nonlinear responses here. The effective phonon bath dimension $d = 1.9 \pm 0.03$ is consistent with the \textmu m-scale wavelength of millikelvin phonons in quartz becoming comparable to the transverse dimensions of the PCR \cite{Behunin2016}.

\begin{table}
\caption{\label{tab:table1} Fit parameters for the data in Fig. \ref{fig:fig2}}
\begin{ruledtabular}
\begin{tabular}{cc}
parameter & value \\
\hline
$F\delta_0^\mathrm{diss}$ & $(1.26\pm0.03)\times 10^{-5}$\\
$\beta$ & $0.56 \pm 0.021$\\
$\overline{n}_c$ & $10 \pm 2.2 $\\
$d$ & $1.9 \pm 0.03$\\
$Q_\mathrm{rel}^{T_0}$ & ($8.3\pm0.43)\times 10^{6}$\\
$T_0$ & $0.25$ K \\
\end{tabular}
\end{ruledtabular}
\end{table}

\section{Identifying TLS sources}
\begin{figure}
\centering
\includegraphics[width=\columnwidth]{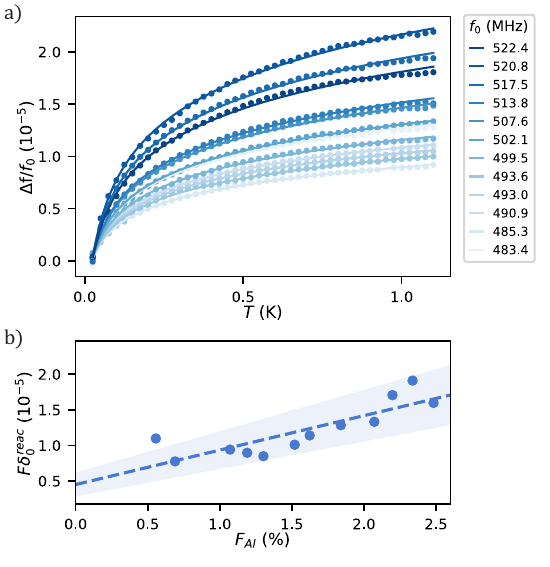}
\caption{Characterizing the TLS response of 12 resonances through reactive measurements (a)  The resonance frequency fractional shift $\Delta f/f_0$ at moderate drive power as a function of temperature. Solid lines are fits to (\ref{eq:resTLS_freqShift}) used to extract $F\delta_0^\mathrm{reac}$. (b) The fitted values of \FReac (blue) plotted against the simulated mechanical participation in the aluminum electrodes $F_{Al}$. A linear fit (blue dashed line) to the data suggests a correlation between the TLS density and mechanical participation in the aluminum. Error bars corresponding to one standard deviation of fit uncertainty in \FReac are smaller than the marker size, and the shaded region corresponds to one standard deviation of uncertainty in the linear fit.}
\label{fig:fig3}    
\end{figure}

Having established that the mechanical dissipation is dominated by TLS coupling, we proceed by characterizing the TLS response of 12 resonances to study the effects of resonator geometry on the bath. To do so, we study the resonator frequency shift corresponding to the reactive change in the mechanical susceptibility. In particular, dispersive interactions with the TLS ensemble induce a frequency shift \cite{Phillips1987}
\begin{equation}\label{eq:resTLS_freqShift}
    \frac{\Delta f}{f_0} = \frac{F\delta_0^\mathrm{reac}}{\pi}\Bigl[\Re \Bigl\{\Psi\Bigl(\frac{1}{2}+\frac{hf_r}{2\pi i k_\mathrm{B}T}\Bigr)\Bigr\}-\ln{\Bigl(\frac{hf_r}{2\pi k_\mathrm{B}T}\Bigr)}\Bigr],
\end{equation}
where $\Psi$ represents the Digamma function, $f_0$ is the resonator frequency at zero temperature, $\delta_0^\mathrm{reac}$ is the reactive TLS loss tangent, F is the filling fraction of TLS in the resonator mode volume, and $\Delta f = f_r(T)-f_0$. Crucially, this expression does not depend on $\overline{n}$ and therefore $f_r(T)$ can be measured with drive powers which are orders of magnitude larger than the TLS saturation power. This results in a signal-to-noise ratio (SNR) which is significantly larger than that corresponding to drive powers used to measure $F\delta_0^\mathrm{reac}$; it is therefore common to use \FReac to rapidly characterize TLS loss \cite{Pappas2011, Gruenke2023}. We emphasize, however, that even tough $F\delta_0^\mathrm{reac}$ and $F\delta_0^\mathrm{diss}$ both characterize the TLS coupling and would be equal for a bath with uniform TLS density and coupling strength, measurements of these two values for the same device need not agree \cite{Phillips1987,Pappas2011}. The potential discrepancy arises from the fact that \FDiss is most strongly influenced by TLSs nearest the resonator frequency, whereas the largest contributions to \FReac are from TLSs for which $hf_\mathrm{TLS}/2k_\mathrm{B}T \gtrsim 1$. The reactive measurement thus characterizes the average TLS density of states and coupling over a large range of TLS frequencies, whereas the dissipative measurement samples a smaller number of TLSs nearby $f_r$ (Appendix \ref{sec:apdx_TLSVariance}). 

We measure \FReac by stepping the base plate temperature of the refrigerator $T$ from 25-1000 mK and monitor $f_r (T)$ for each resonance at moderate power ($100<\overline{n}<400$). We fit the data to (\ref{eq:resTLS_freqShift}) and the fractional frequency shift is plotted in Fig. \ref{fig:fig3}(a). We note that the higher frequency resonances which were designed with larger coupling electrodes exhibit larger $\Delta f/f_0$, a signature of larger $F\delta_0^\mathrm{reac}$. To quantify this effect, we simulate for each resonance the participation of the mechanical energy in the aluminum coupling electrodes, $F_{Al}= E_{Al}/(E_{Al}+E_{qz})$ where $E_{Al}$ and $E_{qz}$ account for the total simulated mechanical energy in the aluminum and quartz volumes of the PCR, respectively. In Fig. \ref{fig:fig3}(b), we plot \FReac against $F_{Al}$ and find that larger values of $F_{Al}$ are correlated with larger $F\delta_0^\mathrm{reac}$. We fit these data to a linear model $F\delta_0^\mathrm{reac}=F_{Al}\delta_{Al}+(1-F_{Al})\delta_{qz}$ to disentangle the quartz and aluminum TLS loss tangents and find $\delta_{qz}$ = $4.5\pm1.6 \times 10^{-6}$ and $\delta_{Al}$ = $4.9\pm1.0 \times 10^{-4}$, respectively. This analysis indicates that the dominant source of TLS defects in these devices is associated with aluminum rather than quartz and further suggests the need to explore new materials or fabrication methods for the electrodes.

\section{Ringdown measurements}\label{sec:Ringdown}
\begin{figure*}
\centering
\includegraphics[width=\textwidth]{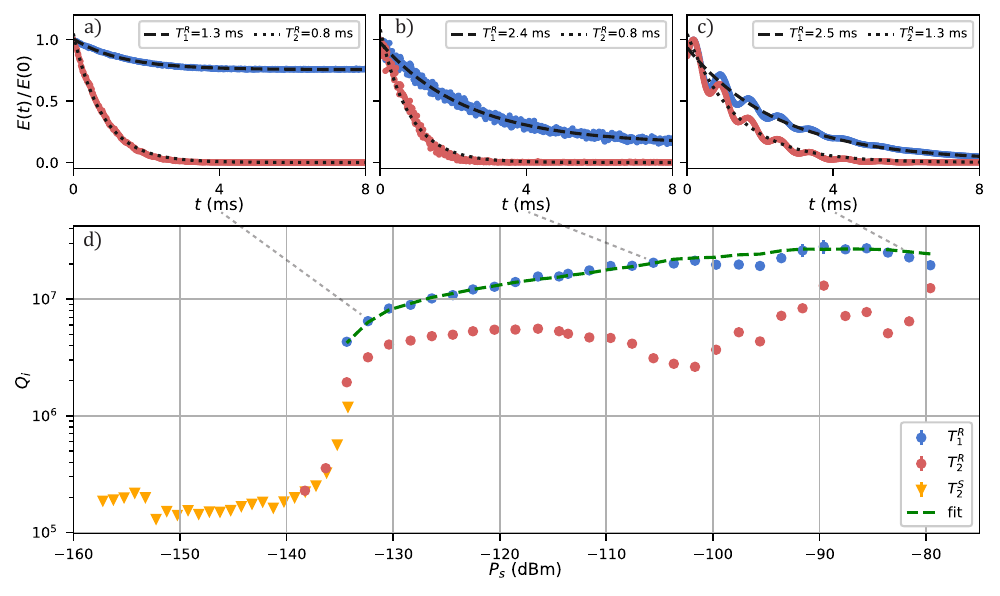}
\caption{Ringdown and spectral measurements of the $f_r = 501.1$ MHz resonance at base plate temperature $T=25$ mK. (a)-(c) Incoherent (blue) and coherent (red) averaging of the resonator energy decay for selected drive powers ($P_s$ = -132.4 dBm, -105.6 dBm, and -81.6 dBm). Exponential fits yield the ringdown decay and decoherence times, $T_1^R$ (dashed) and $T_2^R$ (dotted). (d) The drive power at the sample $P_s$ is swept and $T_1^R$ (blue) and $T_2^R$ (red) are used to calculate $Q_i$. The spectral measurement of the linewidth as measured with a VNA, yielding $T_2^S$, is also related to a quality factor (orange). The loss associated with $T_1^R$ is fit to a model which incorporates resonant TLS damping, radiative leakage, and self-heating-induced relaxation damping (green dashed line).}
\label{fig:fig4}
\end{figure*}

At large drive strengths ($\overline{n} \gtrsim 2000$), dissipated power results in significant self-heating of the PCR owing to the small thermal conductance effected by the phononic bandgap \cite{Cleland2001,Ren2020}. This heats the TLS ensemble, in turn altering the mechanical susceptibility and resulting in a strongly nonlinear response. Furthermore, because this nonlinearity comprises both a reactive and a dissipative shift to the susceptibility, it cannot be modeled with the Duffing formalism \cite{Duffing1921,Kovacic2011, Metzger2024:inPrep}. To probe the decay of the resonators in the nonlinear regime and explore other potential sources of dissipation, we thus turn to time-domain ringdown measurements. 

In this experiment, we study the $f_r=502.1$ MHz resonator by applying a pump tone $f_p$ on resonance for 150 ms and then monitoring the free decay. We choose this resonator as a compromise between maximizing external coupling ($\kappa_e/2\pi=42.2$ Hz, $Q_e= 1.19\times 10^7$) for faster measurement time and minimizing TLS density (\FReac$=1.14\times10^{-5}$) for smaller loss. We average the response (in)coherently over $N$ repetitions of the pulse sequence and fit an exponential model to the data to yield the mechanical coherence (decay) time $T_2^R$ ($T_1^R$) as measured with the ringdown technique. We repeat this measurement over a series of drive powers with the fridge temperature held at $T=25$ mK. In Fig. \ref{fig:fig4}(a)-(c), we present examples of the energy decay at selected $P_s$.

Although the exponential model describes the data well, some ringdown traces appear to have multi-exponential features which may be attributable to small-ensemble TLS relaxation dynamics \cite{Cleland2023}. Furthermore, at very large $P_s$, oscillations in the ringdown energy are observed which we suspect to be an effect of the thermal nonlinearity. We also note that incoherent averaging introduces an offset which reduces the SNR of the $T_1$ measurement as $P_s$ is decreased, limiting our ability to measure $T_1$ at smaller powers. For each fit we disentangle $Q_i$ from the total quality factor $Q=2\pi f_0 T_{1,2}^R$ and we plot in Fig. \ref{fig:fig4}(d) the $Q_i$ associated with $T_{1,2}^R$. We also plot $Q_i$ in the low-power regime as measured with a vector network analyzer (VNA) immediately after the ringdown traces. We denote this VNA measurement of the spectral linewidth as $T_2^S$, and we find that $T_2^R$ and $T_2^S$ are in agreement in the crossover regime. Although we expect for a classical harmonic oscillator that $T_1 = T_2$, at large $P_s$ we find $T_1 > T_2$, suggesting that the device experiences some source of pure dephasing \cite{Maillet2016}. The longest decay time $T_1 = 2.7 \pm 0.05$ ms corresponds to $Q_i = (2.9 \pm 0.20) \times 10^7$ or a quality factor-frequency product $Q_i\cdot f = (1.4 \pm 0.09) \times 10^{16}$. 

We understand this decay to be limited by a combination of resonant TLS damping, relaxation TLS damping, and a power- and temperature- independent loss source which is likely radiative leakage through the finite-numbered mirror cells. Although the relaxation damping should be minimal at $25$ mK, we suspect the device is heated by several tens of mK in this study which exacerbates the relaxation damping. A model which incorporates these effects (Appendix \ref{sec:apdx_ringLossModel}) is fit to the decay data and also plotted in Fig. \ref{fig:fig4}(d). This reaffirms that even at large drive strengths, TLS loss is a dominating source of dissipation at low temperatures and powers. 

\section{Discussion and conclusion}
We have thoroughly investigated loss sources in first-of-their-kind thin-film quartz PCRs at a range of temperatures and powers and found that these devices demonstrate long coherence times. At low phonon number, the measured quality factors represent an order of magnitude improvement over existing piezoelectric PCRs. Although the current device performance is limited in part by coupling to TLS defects in the aluminum or corresponding oxide, measurements of the quartz loss tangent $\delta_0^\mathrm{qz}$ suggest that should this aluminum loss be ameliorated, low power quality factors $Q_i > 200,000$ would be feasible with the current quartz substrate and fabrication methods. Accordingly, future investigation should focus on minimizing the TLS contributions to loss from the coupling electrodes via different materials or fabrication approaches. Little is known, however, about the source of TLS loss in the quartz and this study does not differentiate between surface and bulk contributions. It is also unknown how the top-down manufacturing of thin-film quartz affects its material properties, and other production methods such as epitaxial growth of quartz should be considered. 

The high-power mechanical lifetimes of these devices demonstrate $Q_i\cdot f$-products competitive with bulk acoustic wave resonators on quartz \cite{Galliou2013,Kharel2018} in spite of their much larger surface-to-volume participation. Considering as well their small mass and mode volume, thin-film quartz PCRs may be useful for nanoelectromechanical mass spectrometry \cite{Sage2015,Neumann2024} or other classical sensing applications. We conclude by remarking that the prospects for improving the high-power lifetimes in these devices are strong. Radiative leakage is readily suppressed by extending the number of phononic mirror cell periods. If $F\delta_0$ is further reduced, this would doubly affect the high-power lifetime; not only does reducing $F\delta_0$ directly mitigate dissipation but also it results in less dissipated power and therefore less self-heating-induced relaxation damping.

\begin{acknowledgments}
This material is based upon work supported by the Air Force Office of Scientific Research and the Office of Naval Research under award number FA9550-23-1-0333. We acknowledge support from the Office of the Secretary of Defense via the Vannevar Bush Faculty Fellowship, Award No. N00014-20-1-2833. We thank Zurich Instruments and Edward Kluender for their aid in the setup of the SHFQC+ Qubit Controller for ringdown measurements. We also thank Sarang Mittal, Kazemi Adachi, and Pablo Aramburu Sanchez for fruitful discussions related to this work. 
\end{acknowledgments}

\section*{Data Availability Statement}
The data that support the findings of this study are available from the corresponding author upon reasonable request.

\appendix

\section{Finite element models}
\subsection{Bandgap simulation}\label{sec:apdx_bandgapsim}
The simplest way to generate a phononic bandgap in a 1D elastic waveguide is to structure it in a periodic fashion with a repeating binary pattern. The emergence of a bandgap can be understood by analogy with the toy-model of a 1D chain of diatomic molecules connected with springs \cite{Ashcroft1976}. The dispersion relation features two bands, known in the literature as the \textit{acoustic} and \textit{optical} branches, separated by a frequency gap $\Delta f_b$, the magnitude of which scales linearly with the difference of the two atoms' masses $m_1$ and $m_2$, $\Delta f_b\propto (m_1-m_2)/(m_1+m_2)$. The center frequency $\omega_B$ of the bandgap is determined by the constructive interference condition $\lambda = 2a$, where $a$ is the size of the unit cell in the direction of the waveguide: $f_b=v_\textrm{ph}/2a$, with $v_{ph}$ the phonon velocity in the material. 

In this work we use right-handed ST-cut $\alpha$-quartz, a common orientation of quartz which is valued for its temperature stability \cite{Morgan2010}. In the crystallographic $\hat{x}$ direction of ST-cut quartz -- the axis along which the phononic shields will be patterned -- the longitudinal speed of sound can be approximated $v_l=\sqrt{C_{11}/\rho}\approx 5750$ m/s, where $C_{11}=87.65$ GPa is the elasticity at 5 K and $\rho=2644~\mathrm{kg/m^3}$ is the density \cite{Tarumi2007}.
This suggests that unit cell length $a\approx 5.75$ \textmu m will yield a bandgap with center frequency $f_b=500~$MHz. Higher frequency bandgaps require smaller feature sizes which would in turn increase our sensitivity to fabrication variability; choosing $500$~MHz as our operating frequency appeared to be a good trade-off between fabrication reproducibility and dimensional reduction. Indeed, this guarantees $\lambda/t\approx 10$, so that the thickness dimension $t\approx 1$ \textmu m of the quartz film remains irrelevant. With these design constraints in mind, the fine shape of the unit cell can then be adjusted to optimize the ratio $\Delta f_b/f_b$. 

We simulate the bandgap with finite element method (FEM) simulations in COMSOL \cite{COMSOL} to numerically compute the dispersion of a waveguide with arbitrary geometry and extract the associated bandgap parameters $\{f_b, \Delta f_b\}$. We follow standard conventions for this procedure, performing eigenmode simulations of one mirror unit cell with periodic boundary conditions and sweeping the $k$ vector through the one-dimensional Brillouin zone \cite{Arrangoiz2016}. For the geometric dimensions plotted in Fig. \ref{fig:apdx_defectSim}(a), with $m_x=2.5$ \textmu m, $m_y=3.9$ \textmu m, $b_x=1.0$ \textmu m, and $b_y=1.2$ \textmu m, we predict $f_b=510$ MHz and $\Delta f_b=177$ MHz. Given the lack of commercially-available wideband ($> 200$ MHz) isolators at these frequencies, we do not explicitly study the frequencies over which the bandgap functions but assume that the quality factors we demonstrate would be unachievable without a phononic bandgap.

\subsection{Electromechanical coupling}
In this section we motivate our choice of defect mode by analyzing the strength of the piezoelectric interaction. A mechanical strain $\mathbf{S}(\mathbf{r})$ in a piezoelectric material will induce an electric polarization which interacts with an external electric field $\mathbf{E}$ to produce an interaction described by
\begin{equation}\label{apdx_EMcoupling_piezoHam}
    \mathcal{H}=-\int\mathbf{E}(\mathbf{r})\cdot\mathbf{e}\cdot\mathbf{S}(\mathbf{r})dV,
\end{equation}
where $\mathbf{e}$ is the second-rank piezoelectric coupling tensor in stress-charge form \cite{Jain2023}, $V$ is the defect volume, and the strain is a six-component vector $(S_{xx},S_{yy},S_{zz}, 2S_{yz}, 2S_{xz}, 2S_{xy})^\mathrm{T}$ in Voigt notation. The coupling tensor for ST-cut quartz is given by 
\begin{equation}
    \mathbf{e}=
    \begin{pmatrix}
        0.171 & -0.038 & -0.133 & 0.082 & 0 & 0 \\
        0 & 0 & 0 & 0 & 0.067 & -0.099 \\
        0 & 0 & 0 & 0 & -0.072 & 0.107\\
    \end{pmatrix},
\end{equation}
as expressed in Voigt notation with units of $\mathrm{C/m^2}$ \cite{IEEE1978}. To leverage the relative strength of the largest component $e_{11}$, we design PCRs which couple to a fundamental extensional mode with predominant displacement along $\hat{x}$, thereby maximizing $S_{xx}$. We then design the electrodes to produce an electric field which is also oriented along $\hat{x}$; we idealize this field as $\mathbf{E}=(E_x, 0, 0)^\mathrm{T}$. Considering that the contributions $E_xe_{12}S_{yy}$, $E_xe_{13}S_{zz}$ and $2E_xe_{14}S_{yz}$ are small compared to $E_xe_{11}S_{xx}$ for a fundamental $\hat{x}$-extensional mode, we are left with the interaction
\begin{equation}
    \mathcal{H}=-0.171 E_x S_{xx}V
\end{equation}
by further assuming spatial uniformity of the electric and strain fields. We consider here only fundamental mode excitations to minimize the phononic density of states around the mode of interest. Crucially, although our choice of ST-cut quartz is somewhat arbitrary, crystallographic rotations of $\mathbf{e}$ will not result in a component $e_{ij}>0.171$. To first-order, then, the $\hat{x}$-extensional mode in ST-cut quartz will have as strong of piezoelectric coupling as any fundamental mode in arbitrarily oriented quartz.

\begin{figure}
\centering
\includegraphics{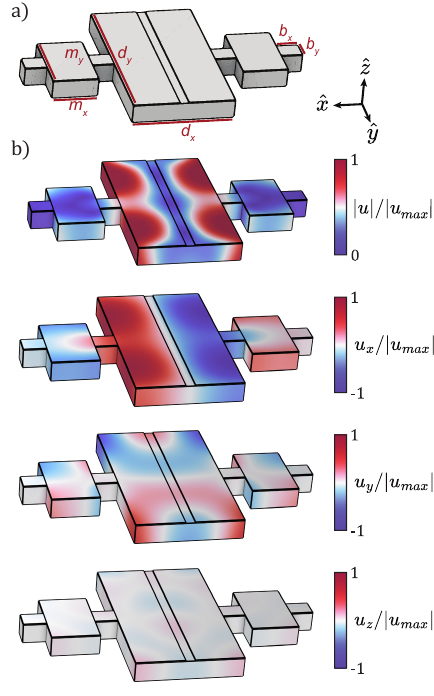}
\caption{\label{fig:apdx_defectSim} Defect mode geometry and polarization. (a) Geometric parameters for a typical defect cell and one period of mirror cells on each side. (b) The defect mode we design coupling to is a fundamental extensional mode with primary displacement along the $\hat{x}$ dimension. In descending order: total displacement of the defect mode, displacement along $\hat{x}$, displacement along $\hat{y}$, and displacement along $\hat{z}$. The color scales are normalized by the maximum total displacement $u_\mathrm{max}$ to illustrate the relative displacements between axes.}
\end{figure}

In Fig. \ref{fig:apdx_defectSim}(b), we plot the simulated displacement of the defect mode and its axial components. This simulation uses an on-resonant frequency-domain voltage drive $V(\omega_r)$ across the coupling electrodes to show the steady-state response for a given drive phase. We typically simulate devices using coefficients of the quartz elastic tensor measured at 5K \cite{Tarumi2007}; we have previously found that these coefficients are highly predictive of quartz mechanical response at low temperatures \cite{Emser2022}. We find that the devices redshift by several MHz as they are cooled from room temperature to mK temperatures and that this is well-modeled by simulations of the devices with both warm and cold elastic tensors.

\subsection{Admittance simulation and decomposition}
We compute the electromechanical admittance $Y(\omega)=I(\omega)/V(\omega)$ of a device by sweeping the frequency $\omega$ of a voltage source across and monitoring the induced current $I(\omega)$ through the electrodes. We then use a vector fitting routine \cite{Gustavsen1999, Arsenovic2022} to approximate $Y(\omega)$ in terms of a rational target function which we ultimately map to an equivalent electrical circuit. Although it would be simple to instead simulate $S_{11}(\omega)$ and fit the Lorentzian response as we do for real data in the main text, the equivalent circuit description not only enables a direct calculation of relevant electromechanical parameters but also facilitates the integration and simulation of the circuit within a larger network. This latter fact is particularly useful for designing hybrid devices with mechanical resonators coupled to superconducting qubits \cite{Arrangoiz2016}. 

For these frequency-domain simulations, we typically prescribe mechanical damping by means of an isotropic structural loss factor $\eta=1/Q_\mathrm{mech}$ where $Q_\mathrm{mech}$ is the desired mechanical quality factor in absence of other dissipation sources. Not only is the added damping useful for more accurately simulating the electrical response of lossy devices, but also a smaller $Q_\mathrm{mech}$ can be helpful for resolving otherwise-sharp resonances in $Y(\omega)$ with a coarser sweep of $\omega$ if $\omega_r$ is not known a priori. This loss necessitates the use of complex poles and residues when fitting $Y(\omega)$ as well as the addition of resistive components in the equivalent electrical circuit.

\begin{figure}
\centering
\includegraphics{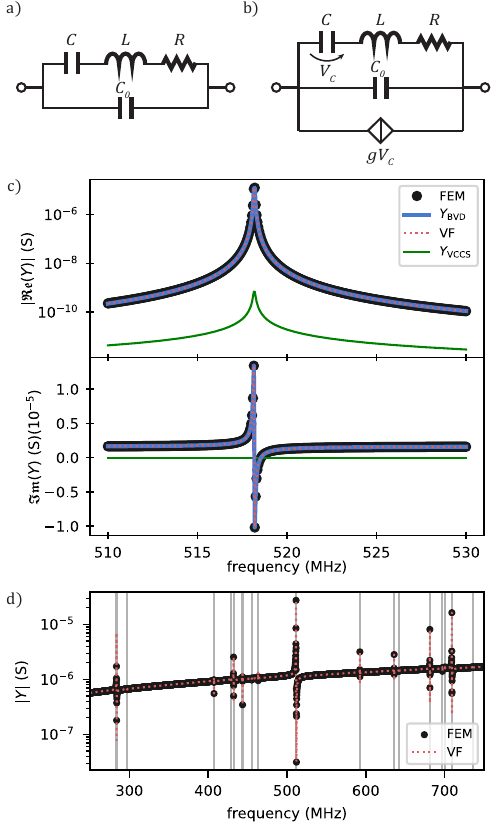}
\caption{\label{fig:apdx_admitSim} Equivalent circuits and admittance simulation. (a) The traditional Butterworth-Van Dyke (BVD) equivalent circuit for modeling a piezoelectric resonance. (b) The equivalent circuit our vector fitting routine maps to, a BVD circuit in parallel with a VCCS. (c) An example of simulated admittance data (black markers) with the full vector fit (red dashed line). We decompose the vector fit to show the contributions from the BVD circuit (blue solid line) and the VCCS (green solid line). The VCCS contribution to $Y(\omega)$ is strictly real and its magnitude is significantly smaller than the BVD contribution. (d) An example of the vector fitting routine simultaneously fitting (red dashed line) 17 pairs of complex poles and residues to the simulated broadband response of a PCR (black markers). Fitted pole frequencies are marked with gray vertical lines.}
\end{figure}

In particular, we seek to describe $Y(\omega)$ with a rational model function consisting of complex poles $\ubar{p}_k=p_k'+ip_k''$ and residues $\ubar{c}_k=c_k'+ic_k''$ which will always come in conjugate pairs $\{\ubar{p}_k, \ubar{p}^*_k\}$ and $\{\ubar{c}_k, \ubar{c}^*_k\}$. We fit in the Laplace domain to the complex target function
\begin{equation}\label{eq:apdx_AdmitSim_targetFunc_1}
    \ubar{F}(s) = \frac{\ubar{c}_k}{s-\ubar{p}_k}+\frac{\ubar{c}^*_k}{s-\ubar{p}^*_k}+es
\end{equation}
where $s=i\omega$ is the complex frequency and $e$ is a proportional coefficient. For reasons which will shortly become clear, we rewrite the target response
\begin{equation}\label{eq:apdx_AdmitSim_targetFunc_2}
    \ubar{F}(s) = \frac{as}{s^2+sc+d}+\frac{b}{s^2+sc+d}+es
\end{equation}
where
\begin{equation}\label{eq:apdx_AdmitSim_targetFunc_3}
    \begin{aligned}
        a&=\ubar{c}_k+\ubar{c}^*_k \\
        b&=-(\ubar{c}_k\ubar{p}^*_k+\ubar{c}^*_k\ubar{p}_k) \\
        c&=-(\ubar{p}_k+\ubar{p}^*_k) \\
        d&=\ubar{p}_k\ubar{p}^*_k. \\
    \end{aligned}
\end{equation}
We would ultimately like to relate $\ubar{F}(s)$ to the Butterworth-Van Dyke (BVD) circuit which is commonly used to model the admittance of a lossy piezoelectric resonator \cite{Butterworth1914, VanDyke1928}. The BVD model consists of a series $RLC$ circuit with resistance $R$, inductance $L$, and capacitance $C$ in parallel with another capacitance $C_0$, as shown in Fig. \ref{fig:apdx_admitSim}(a). Following logic similar to Antonini \cite{Antonini2003}, the admittance $Y_\mathrm{BVD}(s)$ of this circuit is
\begin{equation}\label{eq:apdx_AdmitSim_BVD_1}
    \begin{aligned}
        Y_\mathrm{BVD}(s)&=Y_\mathrm{RLC}(s)+Y_{C_0}(s) \\
        &=\frac{1}{R+sL+1/(sC)}+C_0s \\
        &=\frac{s/L}{s^2+sR/L+1/(LC)}+C_0s,
    \end{aligned}
\end{equation}

which resembles (\ref{eq:apdx_AdmitSim_targetFunc_2}) without the $b$ term. We exploit this similarity by rewriting the target response
\begin{equation}\label{eq:apdx_AdmitSim_targetFunc_4}
    \ubar{F}(s) = Y_\mathrm{BVD}(s)+Y_\mathrm{add}(s)
\end{equation}
where
\begin{equation}\label{eq:apdx_AdmitSim_targetFunc_5}
    Y_\mathrm{add}(s) = \frac{b}{s^2+sc+d}.
\end{equation}
These statements are reconciled if
\begin{equation}\label{eq:apdx_AdmitSim_params}
    \begin{aligned}
        a&= 1/L \\
        c&= R/L \\
        d&= 1/(LC) \\
        e&= C_0.\\
    \end{aligned}
\end{equation}
Unfortunately, this suggests (\ref{eq:apdx_AdmitSim_targetFunc_2}) cannot be represented solely by (\ref{eq:apdx_AdmitSim_BVD_1}) unless $b=-(\ubar{c}_k\ubar{p}^*_k +\ubar{c}^*_k \ubar{p}_k)=0$, and we find no physical reason for this to be generically true. The fitted response is thus described by a BVD circuit which is additionally shunted by an element with admittance $Y_\mathrm{add}(s)$, and the similarity of the denominators in $Y_\mathrm{BVD}(s)$ and $Y_\mathrm{add}(s)$ suggests that this can be realized with the existing topology. Indeed, the desired response is obtained by means of a voltage-controlled current source (VCCS) \cite{Antonini2003}, the control being the voltage $V_{C}$ across the capacitor $C$. This voltage is
\begin{equation}
    \begin{aligned}
        V_C(s) &= \frac{I_\mathrm{RLC}(s)}{sC}=\frac{Y_\mathrm{RLC}(s)V_\mathrm{RLC}(s)}{sC} \\
        &= \frac{V_\mathrm{RLC}(s)}{LC}\frac{1}{s^2+sR/L+1/(LC)}.
    \end{aligned}
\end{equation}
Written in this form, we see that the `missing' admittance is reconciled if 
\begin{equation}
    Y_\mathrm{add}(s)=bLC\frac{V_\mathrm{C}(s)}{V_\mathrm{RLC}(s)},
\end{equation}
which corresponds to a current source with control factor $g=bLC$ where $b$ is specified by (\ref{eq:apdx_AdmitSim_targetFunc_3}). The full circuit including the VCCS is sketched in Fig. \ref{fig:apdx_admitSim}(b). We can finally solve (\ref{eq:apdx_AdmitSim_targetFunc_3}) and (\ref{eq:apdx_AdmitSim_params}) to express the circuit parameters in terms of the fitted poles and residues:
\begin{equation}
    \begin{aligned}
        L &= \frac{1}{\ubar{c}_k+\ubar{c}^*_k} = \frac{1}{2c'_k} \\  
        R &= \frac{\ubar{p}_k+\ubar{p}^*_k}{\ubar{c}_k+\ubar{c}^*_k}= -p'_k/c'_k \\
        C &= \frac{\ubar{c}_k+\ubar{c}^*_k}{\ubar{p}_k\ubar{p}^*_k} = \frac{2c'_k}{|\ubar{p}_k|^2}\\
        C_0 &= e \\
        b &= -(\ubar{c}_k\ubar{p}^*_k+\ubar{c}^*_k\ubar{p}_k) = -2(c'_kp'_k+c''_kp''_k).
    \end{aligned}
\end{equation}

We demonstrate the effectiveness of this routine by simulating $Y(\omega)$ for a standard device and using a vector fitting package \cite{Arsenovic2022} to decompose the response with a pair of complex poles and residues matching (\ref{eq:apdx_AdmitSim_targetFunc_1}). We plot in Fig. \ref{fig:apdx_admitSim}(c) the simulated and fitted admittance response alongside the contributions from the BVD and VCCS parts of the circuit. We find that the device response is well-modeled by only the BVD circuit and that the VCCS contribution is negligible for this and similar devices. The validity of this assumption can be checked by calculating the ratio of admittance contributions for any given fit,
\begin{equation}
    \frac{|Y_\mathrm{VCCS}(s)|}{|Y_\mathrm{BVD}(s)|} = \frac{b}{a|s|}=\frac{1}{\omega}\frac{(\ubar{c}_k\ubar{p}^*_k+\ubar{c}^*_k\ubar{p}_k)}{\ubar{c}_k+\ubar{c}^*_k},
\end{equation}
and ensuring that it is small for frequencies of interest. 

To map the circuit parameters to quality factors, we rewrite the admittance,
\begin{equation}\label{eq:apdx_AdmitSim_YBVDRLC}
     Y_\mathrm{BVD}(s) = sC \frac{\frac{s^2}{\omega_c^2} + \frac{s}{\omega_c Q_i^c}+1+\frac{C_0}{C}}{\frac{s^2}{\omega_p^2}+\frac{s}{\omega_p Q_i^p}+1},
\end{equation}
where $\omega_c=1/\sqrt{LC_0}$ is the residue frequency, $Q_i^c=\sqrt{L/C_0}/R$ is the residue internal quality factor, $\omega_p=1/\sqrt{LC}$ is the pole frequency, and $Q_i^p=\sqrt{L/C}/R$ is the pole internal quality factor; we are interested in the pole parameters. We then find the external coupling rate $\kappa_e$ between the resonator and a transmission line of characteristic impedance $Z_0$ by comparing (\ref{eq:lorentzianFit}) with the S-parameter response of the near-resonant ($\omega\approx\omega_0$) admittance from (\ref{eq:apdx_AdmitSim_YBVDRLC}):
\begin{equation}
    \begin{aligned}
        S &= \frac{\kappa_i - \kappa_e +2i(\omega-\omega_0)}{\kappa_i+\kappa_e+2i(\omega-\omega_0)} \\
        &= \frac{1-Z_0 Y}{1+Z_0 Y} \\
        &\approx \frac{\kappa_i-Z_0/L}{\kappa_i+Z_0/L}.
    \end{aligned}
\end{equation}
We summarize the mapping between circuit elements and resonance theory:
\begin{equation}
    \begin{aligned}
        \omega_r &= \frac{1}{\sqrt{LC}} \\ 
        Q_i &= \sqrt{L/C}/R \\
        Q_e &= \sqrt{L/C}/Z_0 \\
        \kappa_i &= R / L \\
        \kappa_e &= Z_0 / L. \\
    \end{aligned}
\end{equation}
We find that this procedure is capable of fitting the response of many resonances simultaneously by summing the target response over multiple poles:
\begin{equation}\label{eq:apdx_AdmitSim_targetFunc_multiPole}
    \ubar{F}(s) = \sum_k \frac{\ubar{c}_k}{s-\ubar{p}_k}+\frac{\ubar{c}^*_k}{s-\ubar{p}^*_k}+es.
\end{equation}
To demonstrate this, we simulate the admittance of a typical PCR over a broad frequency range, $250-750$ MHz, such that the response of many mechanical modes is visible. We use the vector fitting routine to fit (\ref{eq:apdx_AdmitSim_targetFunc_multiPole}) and plot the result in Fig. \ref{fig:apdx_admitSim}(d); the procedure fits 17 pairs of complex poles and residues and accurately models the simulated response. The robustness and speed of this procedure is particularly useful for iteratively optimizing parameters such as the defect coupling strength.

\section{Radiation leakage}\label{sec:apdx_radLeakage}
\begin{figure}
\centering
\includegraphics[width=\columnwidth]{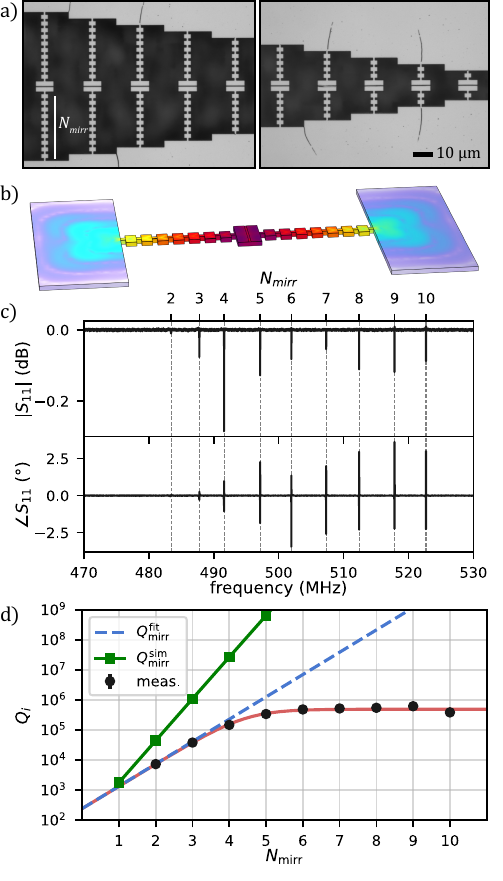}
\caption{\label{fig:apdx_radLeakage} Phononic crystal radiation leakage. (a) SEM images of the device sweeping the number of mirror cells $N_\mathrm{mirr}$. (b) Image of the FEM model used to simulate the radiative loss with the color scale showing the absolute value of displacement plotted on a log scale. (c) Reflected power $S_{11}$ for a probe tone swept across the resonances. We identify nine resonances with $f_r$ consistent with our expectations for the devices $N_\mathrm{mirr}=2-10$. (d) $Q_i$ for each resonator (black markers) is fit to the loss model (\ref{eq:apdx_radLeakageQ}) (red line). The extrapolated radiative leakage (blue dashed line) is much smaller than the simulated radiative leakage (green markers).}
\end{figure}

A separate device was designed to study radiation leakage through the PCR mirror cells. The defect cells in this device maintain a 600 nm electrode gap while sweeping the number of mirror cells $N_\mathrm{mirr}$ between ten and one; the defect area is also swept to step $f_r$ between 475 and 525 MHz. The device is shown in Fig. \ref{fig:apdx_radLeakage}(a). Some features which appear to be tears in the aluminum film can be observed near sharp corners in the geometry. These features -- which appear inconsistently and were not observed in the primary device of this work -- do not seem to be detrimental to device performance. We model the anticipated radiative loss $Q_\mathrm{mirr}^\mathrm{sim}$ with COMSOL as shown in Fig. \ref{fig:apdx_radLeakage}(b). We add large, lossy bounding pads ($Q=10$) to the end of the mirror arrays to force the assumption that any mechanical energy which traverses the mirror cells is dissipated; the bounding pads consist of the same material as the bridge to prevent acoustic reflections. To characterize the actual device, we sweep a high power probe tone ($P_s = -120$ dBm) across the anticipated frequency range and identify nine resonances (Fig. \ref{fig:apdx_radLeakage}(c)) which we map to $N_\mathrm{mirr}=2-10$ and assume that mode $N_\mathrm{mirr}=1$ is too lossy to resolve. 

With moderate drive power $P_s=-140$ dBm, we measure and fit further traces for each resonance and in Fig. \ref{fig:apdx_radLeakage}(d) we plot the resulting values of $Q_i$. We find that $Q_i$ increases exponentially for small $N_\mathrm{mirr}$ and then plateaus around $N_\mathrm{mirr}=6$. Suspecting that the high $N_\mathrm{mirr}$ loss is dominated by resonant TLS dissipation, we fit the data to the loss model
\begin{equation}\label{eq:apdx_radLeakageQ}
    Q_{i}^{-1} = (Q_\mathrm{mirr}^0)^{-1}e^{-\beta N_\mathrm{mirr}} + Q_\mathrm{TLS}^{-1},
\end{equation}
which assumes exponential suppression of radiative leakage with $N_\mathrm{mirr}$. We find $Q_\mathrm{mirr}^0=(4.1\pm0.1)\times10^{-3}, \beta=1.71\pm0.02$, and $Q_\mathrm{TLS}=(4.9\pm0.33)\times10^5$. For $N_\mathrm{mirr}=7$, the number of mirrors in the primary device studied in this work, we expect a radiation-limited quality factor of $3.85\times10^7$.

Although this gives us a sense of the radiative leakage, there are some limitations to this study. Foremost, because we apply a constant drive power which results in varying $\overline{n}$ and because the TLS ensemble is different for each resonator, the parameter $Q_\mathrm{TLS}^{-1}$ is an oversimplification of the TLS dissipation. This means that we are likely underestimating the TLS dissipation for the low $N_\mathrm{mirr}$ devices which most heavily weight $\beta$. This could be avoided by using high-power ringdown measurements to suppress resonant TLS interactions, a measurement technique that we had not yet explored at the time we measured this device. Furthermore, this device was fabricated on a separate chip from and cooled down in a different refrigerator than the device studied in the main text. Although the fabrication methods were similar, it is possible that the radiative leakage varies from chip to chip. 

\section{Ringdown measurements}\label{sec:apdx_ringLossModel}
\subsection{Ringdown technique}
The readout pulses used in the ringdown measurements were generated and analyzed using the double superheterodyne up- and down-conversion stages of a SHFQC synthesizer from Zurich Instruments. Because of the relatively high mechanical $Q\lesssim 20\times 10^6$ at high drive power, it may take as long as $3 Q/f_r\sim120~$ms for the resonator field to settle. We therefore use $150~$ms long squares pulses (with a $150~$ms delay between pulses) to ensure that we are probing the steady-state mechanical response. Fig.~\ref{fig:apdx_fig4theory}(a) shows the time-domain response to such a pulse after demodulation.

Ringdown measurements are performed by sending $N$ such readout pulses at the resonator frequency and recording for each pulse the demodulated in-phase $I(t)$ and quadrature $Q(t)$ time responses. To perform incoherent averaging we then compute $E_\textrm{mag}(t) = \langle A(t)^2\rangle_N = \langle I(t)^2 + Q(t)^2\rangle_N$. Because this method of averaging discards the phase information it is insensitive to frequency fluctuations and thus yields the energy relaxation time $T_1^R$. Alternatively, a coherent averaging scheme $E_\textrm{cpx}(t) = \langle I(t)\rangle_N^2 + \langle Q(t)\rangle_N^2$ allows one to access the mechanical $T_2^R$ decay time. This scheme preserves phase information and is therefore sensitive to fluctuations of the resonance frequency \cite{Maillet2016}. Coherent (complex) averaging is also known to yield a $\sqrt{N}$-times better SNR than incoherent (magnitude) averaging, even for weak signals and a small number of averaged traces \cite{Baumann2019}. In absence of dephasing, one can show that $E_\textrm{mag} = E_\textrm{cpx} + \sigma_I^2 + \sigma_Q^2$, where $\sigma_X^2=\langle X^2\rangle_N - \langle X\rangle_N^2$, which explains the amplitude offset in the magnitude-averaged ringdowns used to extract $T_1^R$.

\subsection{Ringdown loss model}
In this section we demonstrate that the power-dependence of the internal loss extracted from the ringdown measurements presented in the main text can be fit by combining the standard TLS tunneling model with a model for self-heating from dissipated power in the defect resonator. To do so, we find $f_r$ from characteristics of the reflected ringdown signal which we then use to infer an effective resonator temperature $T_\mathrm{eff}$ based on previous measurements of $f_r(T)$ as well as the intracavity phonon number \nbar. This amounts to using the frequency of the PCR as a thermometer for the TLS bath to which it is coupled.

\begin{figure*}
\centering
\includegraphics[width=\textwidth]{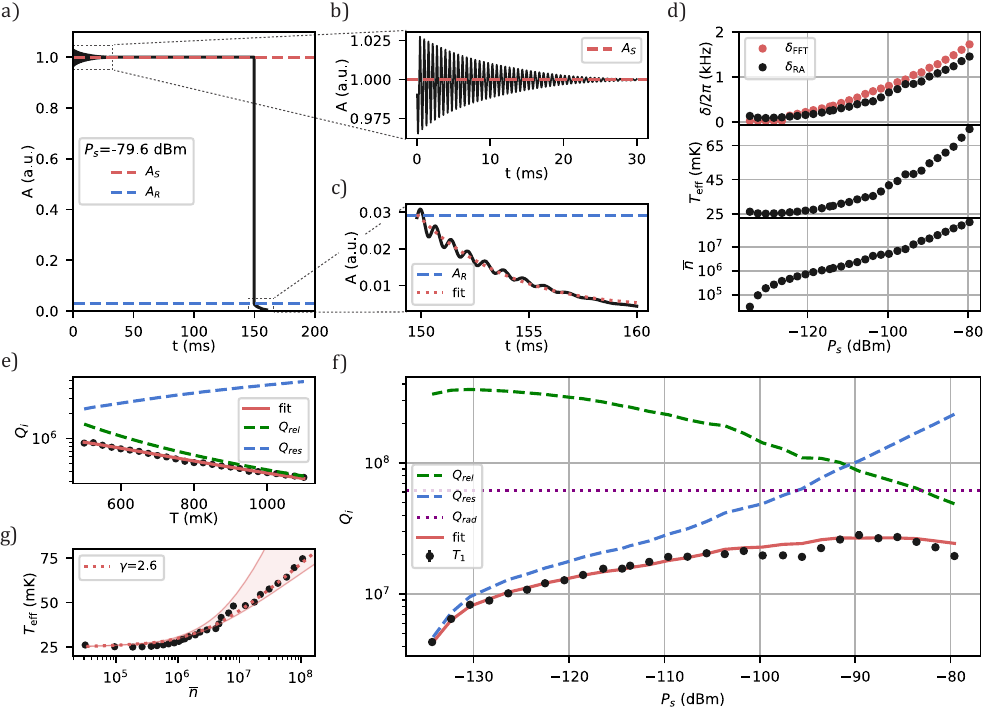}
\caption{Ringdown loss and self-heating in the 502.1 MHz resonance. (a) The measured ringdown amplitude including the reflected signal with the pump tone on. (b) The promptly reflected signal showing oscillations from the beating of $f_{p}$ and $f_r$. The signal decays to (normalized) level $A_s$. (c) The resonator decay after the pump tone is turned off. (d) From top to bottom: the resonator detuning $\delta$ from the pump tone as calculated with both ringdown-amplitude (black) and Fourier transform (red) methods, the effective temperature $T_\mathrm{eff}$ extrapolated from the detuning, and the intracavity phonon population. (e) The data and fit used to find $Q_\mathrm{rel}^{T_0}$. The fit is broken into relaxation (green dashed line) and resonant (blue dashed line) contributions. (f) The ringdown decay internal quality factors are fit to find the resonant TLS dissipation (dashed blue lines) and power- and temperature-independent background loss (purple dotted line) while the relaxation damping (green dashed line) is inferred from $T_\mathrm{eff}$. We assume the background loss $Q_{rad}^{-1}$ is dominated by radiative phonon leakage through the mirror cells. (g) A simple model for the thermal conductance of the phononic mirror cells is used to fit $T_\mathrm{eff}(\overline{n})$. The fit yields temperature exponent $\gamma=2.6\pm0.2$, shown with a red dotted line. The shaded region shows the range of $T_\mathrm{eff}$ that would be realized with $\gamma=1$ (upper bound) to $\gamma=3$ (lower bound). }
\label{fig:apdx_fig4theory}
\end{figure*}

We first review the expected output signal $A_\mathrm{out}(t)$ after a drive pulse with a step excitation is incident on a cavity with total loss rate $\kappa=\kappa_e + \kappa_i$. We model the drive pulse used for ringdown measurements as a step excitation with amplitude $A_p$ and angular frequency $\omega_p$,
\begin{equation}\label{eq:apdx_rngTheory_VIn}
    A_\mathrm{in}(t)=A_p e^{i\omega_p t}\cdot\Theta(t),
\end{equation}
and we assume that the mechanical resonator at $\omega_r=2\pi f_r$ is detuned from $\omega_p$ by $\delta=\omega_p-\omega_r$. The time evolution of the outgoing pulse is obtained by Fourier transform \cite{Metzger2022},
\begin{equation}\label{eq:apdx_rngTheory_VOut_Fourier}
    A_\mathrm{out}(t)=\mathcal{F}^{-1}[\mathcal{F}[A_\mathrm{in}(t)] \cdot S_\mathrm{11}(\omega)]
\end{equation}
which evaluates to
\begin{equation}\label{eq:apdx_rngTheory_VOut}
    A_\mathrm{out}(t)=A_p e^{i\omega_p t}[1-\frac{2\kappa_e/\kappa}{1+2i\delta/\kappa}(1-e^{-\kappa t/2}e^{-i\delta t})]\cdot\Theta(t).
\end{equation}
At short time scales, we thus expect the promptly reflected signal to oscillate at the pump-resonator detuning frequency $\delta$ before stabilizing in the steady-state limit to $A_{S}=A_\mathrm{out}(\infty)e^{-i\omega_p t}=A_p(1-\frac{2\kappa_e/\kappa}{1+2i\delta/\kappa})$, after demodulation at $\omega_p$. This transient oscillation can be understood as the beating between the drive field at angular frequency $\omega_p$ and the impulse response ringing down at the resonator frequency $\omega_r$.
Once the drive pulse is turned off, energy will begin leaking out of the cavity with initial amplitude $A_R=A_p(\frac{2\kappa_e/\kappa}{1+2i\delta/\kappa})$. We eliminate the drive amplitude dependence by taking the ratio of the steady-state and initial ringdown amplitudes,
\begin{equation}\label{eq:apdx_rngTheory_ampRatio}
    \frac{|A_R|}{|A_S|} = \frac{2\kappa_e/\kappa}{\sqrt{(1-2\kappa_e/\kappa)^2+4(\delta/\kappa)^2}},
\end{equation}
which we rewrite as 
\begin{equation}\label{eq:apdx_rngTheory_detuningExpr}
    \delta \approx \pm\kappa_e\sqrt{\left(\frac{|A_S|}{|A_R|}\right)^2-\left(\frac{\kappa}{2\kappa_e}-1\right)^2},
\end{equation}
where the sign is deduced from whether $\omega_p$ is initially chosen $>\omega_r$ (+) or $<\omega_r$ (-) at vanishing power $P_s$, and we approximated the power-dependent resonance frequency by its bare value $\omega_r$.

In Fig. \ref{fig:apdx_fig4theory}(a), we plot the amplitude of the output ringdown signal for a high power measurement ($P_s$ = -79.6 dBm) with a 150 ms drive pulse. The data are scaled such that $A_{S}=1$. We zoom in on the early time dynamics of the promptly-reflected signal in Fig. \ref{fig:apdx_fig4theory}(b) for which oscillations consistent with (\ref{eq:apdx_rngTheory_VOut}) suggest a significant detuning between the pump and resonator frequencies. In Fig. \ref{fig:apdx_fig4theory}(c) we show the resonator decay after the drive tone is turned off which we fit as in the main text. With these data we calculate the pump-resonator detuning $\delta$ in two ways: by Fourier analysis of the promptly reflected signal ($\delta_\mathrm{FFT}$) and by comparing the relative amplitudes the steady-state and ringdown signals ($\delta_\mathrm{RA}$). We perform this analysis for ringdown traces at every sample power and plot the detunings in the top row of Fig. \ref{fig:apdx_fig4theory}(d); we find that $\delta$ increases monotonically with $P_s$ in both analyses. For small powers $P_s<-125$ dBm, a large DC offset in the Fourier transform of the promptly reflected signal obfuscates the detuning signal and we assume $\delta_\mathrm{FFT}$=0. The relative amplitude detuning does not suffer this problem, and so we assume $\delta=\delta_\mathrm{RA}$ below.

We note that the previous analysis neglects the time dynamics of the readout power heating and assumes that the resonator instantaneously reaches its steady-state temperature when the drive is turned on. This assumption is confirmed by the fact that the measured initial transients decay with a fixed frequency, as seen in Fig. \ref{fig:apdx_fig4theory}(b). If the resonator were still thermalizing during the initial transients, then its resonance frequency would shift accordingly and the transients would show some chirp. The measured time responses therefore suggest that the thermalization dynamics are faster than $\approx 1$ ms.

Having calculated the pump-resonator detuning, we proceed by using the characterization of $f_r(T)$ in the main text Fig. \ref{fig:fig3}(a) to infer the effective temperature \Teff of the resonator based on the detuning frequency: $\omega_r(T) = \omega_p + \delta(T)$. We assume that the detuning is positive-valued and that the resonator is well-thermalized to the mixing chamber temperature at small drive powers, $T_\mathrm{eff}=T_\mathrm{MXC}=25$ mK. We plot \Teff in the middle row of Fig. \ref{fig:apdx_fig4theory}(d) and show that the resonator is heated by over 40 mK with large pump powers. We also calculate the steady-state intracavity phonon number assuming that the pump is detuned from the resonance \cite{Aspelmeyer2014},
\begin{equation}\label{eq:apdx_rngTheory_nbarDetuned}
    \overline{n} = \frac{\kappa_e}{\delta^2+(\kappa/2)^2}\frac{P_s}{\hbar\omega_r},
\end{equation}
which we plot in the bottom row of Fig. \ref{fig:apdx_fig4theory}(d). We note that although $P_s$ is swept over six orders of magnitude, \nbar varies by only three orders of magnitude due to the monotonically increasing $\delta$.

We anticipate three significant dissipation mechanisms in this regime of power and temperature: resonant TLS damping, relaxation TLS damping, and radiation leakage through the finite-numbered mirror cells. We can study the relaxation damping expected in this resonator by using the dataset from Fig. \ref{fig:fig3}(a) in the main text but this time studying $Q_i(T)$ instead of $f_r(T)$. With these data, we fit the relaxation damping contribution using the joint loss model (\ref{eq:jointQModel}) and plot the results in Fig. \ref{fig:apdx_fig4theory}(e). We find $d=1.84\pm0.10$ and $Q_\mathrm{rel}^{T_0}=(1.48\pm0.12)\times10^6$ for reference temperature $T_0=0.5$ K. With these parameters fixed, we then use \Teff to calculate the relaxation damping contribution to dissipation $Q_\mathrm{rel}^{-1}(T_\mathrm{eff})$ and fit the ringdown measurement series to find the remaining resonant and background contributions. To better constrain the resonant TLS damping model, we assume \FDiss$=1\times 10^{-5}$. The result is plotted in Fig. \ref{fig:apdx_fig4theory}(f) with the individual loss contributions from resonant, relaxation and background losses. Although we do not anticipate that the resonant TLS loss model should apply at such large powers, we find reasonable fit parameters $\overline{n}_c=10.3\pm3.7$ and $\beta=0.84\pm0.03$. The power- and temperature-independent background loss, which we assume to be radiation leakage through the mirror cells, is found to be $Q_\mathrm{rad}=(6.14\pm0.74)\times10^7$. This value is consistent with our expectation of radiative leakage for $N_\mathrm{mirr}=7$ from Appendix \ref{sec:apdx_radLeakage}.

We find that the highest quality factors are obtained when the three contributions to loss are nearly equal. The first lesson from this study is clear: future devices should be fabricated with larger $N_\mathrm{mirr}$ to suppress the radiative leakage. It is more difficult, however, to suppress the resonant and self-heating relaxation loss contributions. At the largest powers ($P_s > -85$ dBm), $Q_i$ decreases with increasing drive strength as more dissipated power results in more self-heating and subsequently more relaxation damping. This turnover may be most obviously improved by decreasing \FDiss and $F\delta_0^\mathrm{reac}$, which would not only decrease the resonant damping associated with a given $\overline{n}$, but also decrease the dissipated power and subsequent self-heating.

\subsection{Thermal conductance model}

To demonstrate that the apparent self-heating is consistent with microwave power dissipated in the device, we now show that the readout power dependence of the inferred defect site temperature $T_\mathrm{eff}$ can be accounted for by a simple thermal conductance model.
We assume microwave absorption as the only source of heating and denote as $T_0$ the local temperature of the defect site in absence of any microwave drive. Under steady state conditions, the power flow into the hot defect bath due to microwave absorption, $P_\mathrm{in}$, equals the power flow from the hot defect into the chip bath, $P_\mathrm{out}$, and the defect thermalizes at an effective temperature $T_\mathrm{eff}$.
The microwave power dissipated in the defect can be expressed in terms of the average phonon number $\overline{n}$,
\begin{equation}\label{eq:apdx_rngTheory_powerDiss}
    P_\mathrm{in}=\overline{n}\hbar\omega_r^2/Q_i.
\end{equation}
The hot defect bath loses energy via coupling to lattice phonons, which radiate into the chip bath through the two patterned beams that mechanically support the defect site.
The rate at which heat is removed from the defect can be expressed in terms of the lattice thermal conductance $G_{th}(T)$ of the two beams,
\begin{equation}\label{eq:apdx_rngTheory_powerOut}
    P_{out}=\int_{T_0}^{T_\mathrm{eff}} G_{th}(T)d T. 
\end{equation}

For the temperature range considered in this work, the lattice thermal conductance should scale as a power law of the phonon bath temperature $G_{th}\propto T^\gamma$, with the temperature exponent $\gamma$ depending on the effective phonon dimensionality.
When $T\ll \hbar v_{ph}/(k_B w)\approx 10$~mK, the dominant phonon wavelength becomes larger than the beam typical transverse dimension, $w$, and the latter therefore behaves as an effective one-dimensional waveguide for lattice phonons. In this regime, $\gamma=1$ and one expects $G_{th}\propto T$ with the quantized universal coefficient $g_0/T = \pi^2 k_B^2/(3 h)$ \cite{Rego1998}. Each of the two beams should support four populated phonon modes, thus we expect that $G_{th}$ should approach $8 g_0$ in the case of ideal coupling between the beams and the rest of the chip \cite{Schwab2000}. As temperature is increased, however, the thermal conductance experiences a crossover from one-dimensional to three-dimensional behavior, and the Debye law corresponding to $\gamma=3$ is eventually recovered \cite{Wang2007}. At intermediate temperatures for which the thermal wavelength is comparable to the phononic crystal unit cell length, a suppression of the thermal conductance is expected \cite{Cleland2001}.

For the ringdown data shown in Fig. \ref{fig:apdx_fig4theory}(d), $T_0\approx 25$~mK and the device thermal conductance is not expected to be quantized. Although $T_\mathrm{eff}$ spans a range $T_0$ to $\sim 3T_0$ over which the phonon dimensionality will vary, we make the simplifying assumption of a fixed $\gamma$ which yields $G_{th}(T) = G_{th}(T_0)(T/T_0)^\gamma$. Plugging this expression into (\ref{eq:apdx_rngTheory_powerOut}) and solving for $T$, we obtain
\begin{equation}\label{eq:apdx_rngTheory_TeffPdiss}
    T_\mathrm{eff} = T_0\left(1 + \frac{1+\gamma}{T_0\, G_{th}(T_0)}P_{in}\right)^\frac{1}{1+\gamma}.
\end{equation}
In the limit that $P_{in}\ll G_{th}(T_0)T_0$, the readout power heating remains small and this equation reduces to $T_\mathrm{eff}\approx T_0 + P_{in}/G_{th}(T_0)$. By combining (\ref{eq:apdx_rngTheory_powerDiss}) and (\ref{eq:apdx_rngTheory_TeffPdiss}), we fit $T_\mathrm{eff}(\overline{n})$ with $\gamma$ and $G_{th}(T_0)$ as the only fitting parameters. The best fit, shown in Fig. \ref{fig:apdx_fig4theory}(g), yields $\gamma = 2.6 \pm 0.2$ and a thermal conductance $N=1.6\pm 0.3$ as expressed in terms of $G_{th}=N g_0(T_0)$ for $T_0=25$~mK. The fitted value of $\gamma$ agrees with numerical simulations of the beam thermal conductance which follow the procedure described in Ref. \cite{Cleland2001}; we predict $2.5<\gamma(T)<2.8$ with $25<T<75$~mK. The extracted number of conductance channels $N$ being smaller than the expected eight may suggest that the phonon modes of the beam are not perfectly transmitted to the cold bath; this is consistent with the fact that the beam geometry was not optimized for adiabatic coupling with the chip reservoir \cite{Rego1998}. We emphasize that although we oversimplify the model by assuming constant phonon dimensionality and a posteriori extrapolate a quantized number of thermal conductance channels despite $T_0$ being slightly greater than the limit for which we expect this to be valid, the rough agreement of this simple model with the measured data suggests that this picture for thermal conductance and device heating from dissipated power is notionally accurate.

\section{Fabrication}\label{sec:apdx_fabrication}
Fabrication begins with a wafer of diameter 100 mm sourced from NGK Electronics which is comprised of 1 \textmu m, right-handed ST-cut $\alpha$-quartz bonded to 500 \textmu m of Float-zone silicon with a 30 nm amorphous silicon bonding layer. The wafer was produced with a top-down manufacturing process; a 500 \textmu m quartz wafer was bonded to silicon and the quartz was then lapped and polished to the desired thickness. Ellipsometry measurements show that the quartz thickness is 1014 $\pm$ 9 nm over the entire 100 mm wafer.  

The wafer is first cleaned with heated Nano-Strip\textregistered~ solution for 10 minutes at 80 \textdegree C. A grid of alignment marks Ti (10 nm) / Au (100 nm) is then evaporated and lifted off in preparation for chip-scale electron-beam lithography (EBL). Next, the wafer is diced into 10 mm $\times$ 10 mm chips and an aluminum hardmask is patterned to prepare the chip for etching through the quartz. A 250 nm layer of aluminum is evaporated onto the chip (Angstrom Quantum series evaporator) and then patterned with EBL.  All EBL is performed with a JEOL JBX-6300FS. The aluminum is then etched with a $\mathrm{Cl_2+BCl_3+Ar}$ plasma (Oxford Cl ICP etcher) and the resist is stripped to complete the hardmask. The quartz is then fully etched through at the exposed regions of the hardmask with a $\mathrm{CHF_3+O_2}$ plasma (Oxford Fl ICP etcher) etching at an approximate rate of 85 nm/min. The chip is cleaned with an $\mathrm{O_2}$ plasma (Technics PE-II Asher) and the aluminum hardmask is removed with Transene etchant type A. Resist for EBL is then spun and subsequently patterned to define the aluminum coupling electrodes and CPW for each device via liftoff. After exposure and development, the chip is treated with an $\mathrm{O_2}$ descum process before 30 nm of aluminum is evaporated at an approximate rate 1 Å/s. The aluminum features are lifted off in a solution of heated Remover PG at 70 \textdegree C. A secondary dicing step then reduces the chip size to 6 mm $\times$ 6 mm to remove the area of the chip which experienced significant edge beading during the previous processing steps. Finally, a $\mathrm{XeF_2}$ process with a helium carrier gas etches the silicon and suspends the PCRs. The etch is conducted in cycles with an etch rate of approximately 1 \textmu m per cycle after oxide breakthrough.

\section{Measurement and wiring}\label{sec:apdx_wiring}
\begin{figure}
\centering
\includegraphics[width=\columnwidth]{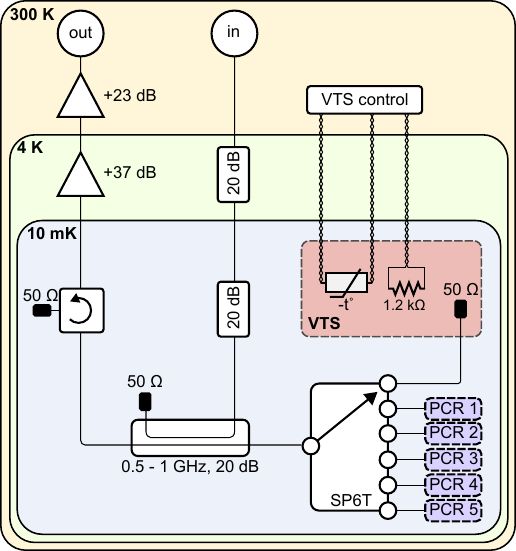}
\caption{Wiring diagram for PCR measurement and gain calibration. For device characterization, a Keysight VNA is used to probe the microwave reflection off the PCR devices through the `in' and `out' ports. A cryogenic six-port RF switch (SP6T) allows us to separately probe five different PCR devices, each with multiplexed arrays of resonators. To perform gain and attenuation calibration we switch the SP6T to connect to the $50~\Omega$ load on the variable temperature stage (VTS). The VTS temperature is adjusted by using a PID control loop set by a Lakeshore 340 temperature controller to regulate the current flowing through the VTS heater. We monitor the VTS temperature with a RuOx sensor thermally connected to the $50~\Omega$ load.}
\label{fig:apdx_wiring}
\end{figure}

All measurements are performed in an Oxford Instruments Triton\texttrademark\ Cryofree dilution refrigerator. The microwave and DC wiring used in this work are schematized in Fig.~\ref{fig:apdx_wiring}. The measured chip has a total 12 CPWs which are shunted by multiplexed arrays of PCRs; we report on only two such devices in this work. We use a cryogenic SP6T coaxial RF switch (Radiall R583423141) to probe up to five of these devices in the same cooldown without needing dedicated input/output lines for each device. The sixth RF switch port is connected to a $50~\Omega$ termination which is used as a broadband noise source to calibrate the output line’s total gain (in Appendix~\ref{sec:apdx_VTScalib}).

We use a 0.5-1~GHz 20dB directional coupler (Pasternack PE2200-20) to couple the input microwave signal to the device. The output line is comprised of a $500$~MHz cryogenic low-noise amplifier (Berkshire U-500-2, 1.1~K noise temperature and 37~dB of gain at $500$~MHz) mounted on the $4$K plate of the dilution refrigerator and a $450-550~$MHz Raditek isolator to protect the devices from the amplifier noise. At room temperature, the output signal is further amplified with a 0.4-3~GHz amplifier (Mini-Circuits ZX60-P33ULN+). The reflection measurements are taken with a Keysight E5071C vector network analyzer.

\section{Calibrated gain and attenuation}\label{sec:apdx_VTScalib}
To accurately calculate $\overline{n}$, one must know the attenuation between the input port and sample to relate some input microwave power $P_\textrm{in}$ to the power at the sample $P_s$. Although one could rely on a room-temperature measurement of the transmission through the dilution refrigerator to the sample prior to cooling, the RF cable attenuation will decrease upon cooling which would result in an underestimation of $\overline{n}$. A careful calibration of the input line attenuation at cryogenic temperature is therefore required. We accomplish this by calculating the gain from the sample with the Y-factor method \cite{Keysight2018, Fernandez1998} and then subtracting the gain from the total transmission through the measurement circuit to yield the attenuation to the sample.

\begin{figure*}
\centering
\includegraphics[width=\textwidth]{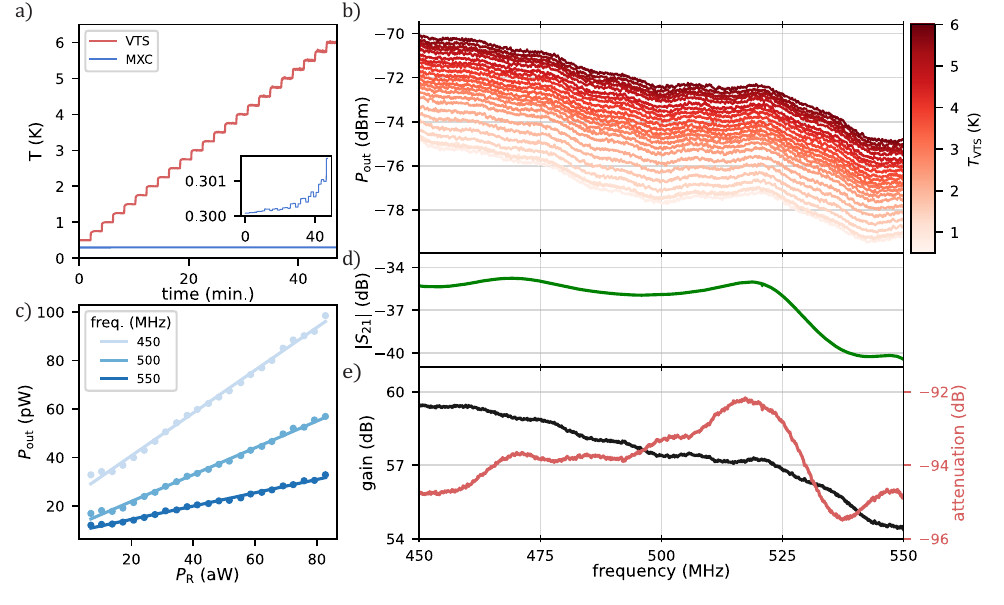}
\caption{ Gain and attenuation calibration with a variable temperature stage (VTS). (a) Measured temperatures of the VTS (red) and the MXC (blue) as a function of time. Starting from $T_i=500~$mK, the PID target temperature for the VTS is increased by $250~$mK until the final temperature $T_f=6~$K is reached. While $T_\mathrm{VTS}$ is stepped, $T_\mathrm{MXC}$ is held at $300~$mK with a separate PID loop. Inset: a zoom-in on the $T_\mathrm{MXC}$ data shows that slight heating of the MXC ($<1.5$ mK) is observed at the highest VTS temperatures. (b) Output power $P_\textrm{out}$ as a function of frequency measured at each temperature step with an Agilent E4407B spectrum analyzer. (c) $P_\mathrm{out}$ as a function of the Johnson–Nyquist noise power $P_\textrm{R}$ from the $50~\Omega$ termination on the VTS for select frequencies (shaded blue markers). The gain is equivalent to the slope of a line fit to the data for each frequency (shaded blue lines). (d) The magnitude of total transmission $|S_{21}|$ from input to output. (e) We plot the gain $G(f)$ from the $50~\Omega$ termination to the measurement output for each sampled frequency $f$ (black line). The attenuation to the sample (red line, right axis) is calculated from the difference of the transmission and gain measurements.}
\label{fig:apdx_gainCal}
\end{figure*}

We measure the output line gain by means of a variable-temperature stage (VTS). The VTS consists of a $\sim 1"\times 1"$ copper plate on which we mount a $50~\Omega$ load, a $1.2~\mathrm{k\Omega}$ surface-mount resistive heater, and a calibrated RuOx thermometer. The VTS is then anchored to the mixing chamber (MXC) plate of the dilution refrigerator via a stainless steel bracket. The weak thermal contact of the bracket allows us to vary the temperature $T_\textrm{VTS}$ of the VTS -- hence also the emitted microwave power from the $50~\Omega$ load -- without significantly affecting the mixing chamber temperature $T_\mathrm{MXC}$. We monitor $T_\mathrm{VTS}$ and control the current through the resistive heater with a Lakeshore Model 340 temperature controller. As shown in Fig. \ref{fig:apdx_wiring}, we use a six-port cryogenic RF switch (Radiall R583423141) to switch the input/output lines between the $50~\Omega$ load and the PCR devices. The cable lengths from the switch to the PCR devices and from the switch to the $50~\Omega$ load are equivalent. 

The thermal noise produced by a resistor can be modeled by its Thévenin equivalent circuit: a voltage source (with mean square voltage $\overline{v_n^2}$) representing the noise of the non-ideal resistor in series with an ideal noise-free resistor $R_0$. The Johnson–Nyquist noise power spectral density produced by a resistor $R$ at temperature $T$ and frequency $f$ is given by
\begin{align}
    S_\textrm{R}&= \frac{\overline{v_n^2}}{\Delta F} =4 h f R\Bigl( \frac{1}{2}+\frac{1}{e^{\frac{h f}{k_B T}}-1} \Bigr)\nonumber\\
    &= 2 h f R \coth{\Bigl(\frac{hf}{2 k_B T}\Bigr)}\nonumber\\
    &\underset{\mathclap{k_B T\gg h f}}{\approx} 4\, R \,k_B T,
\end{align}
where the $1/2$ contribution in the first line accounts for quantum fluctuations \cite{Callen1951} and $\Delta F$ is the bandwidth over which the noise is measured. The dissipated power from the resistor then reads
\begin{equation}
    P_R=\frac{\overline{v_n^2}}{R} = \frac{S_R \Delta F}{R} = 4\, \Delta F\,k_B T.
\end{equation}
When embedding the resistor in a circuit, \textit{i.e.} when connecting it to a load $R_l$, the noise voltage drop across the load is $v_l=R_l/(R_l+R_0)v_n=v_n/2$ when both resistors are matched to $50~\Omega$. The resulting noise power at the load is therefore reduced by a factor 4, such that $P_R=\Delta F\,k_B T$. The output noise measured by the spectrum analyzer is related to the emitted noise power by $P_\textrm{out}(f)=G(f)[h f \Delta F N_\textrm{sys}(f) + P_R(f)]$, with $G(f)$ the net gain of the output line and $N_\textrm{sys}(f)$ the effective noise quanta at frequency $f$ added to the signal at the output port due to amplifiers and losses. 

As shown in Fig. \ref{fig:apdx_gainCal}(a)-(b), we vary $T_\textrm{VTS}$ from $0.5$ to $6$ K and measure the thermal noise at the output of the amplification chain with an Agilent E4407B spectrum analyzer. We average with 1~MHz resolution bandwidth and use PID control for both the VTS and the MXC which we hold at $T_\mathrm{MXC}=0.3$ K to minimize heating from the VTS. Fitting a line to the measured $P_\textrm{out}$ \textit{vs.} $P_\textrm{R}$ data measured at various VTS temperatures, one can extract the gain $G$ from the slope, as illustrated in Fig.~\ref{fig:apdx_gainCal}(c). In Fig.~\ref{fig:apdx_gainCal}(d), we show the total transmission $|S_{21}|$ measured through the fridge. The output gain in our frequency range of interest, $f\in[450,550]~$MHz is shown in Fig.~\ref{fig:apdx_gainCal}(e). At 500~MHz, we extract $G=57.4~$dB, which is consistent with the nominal gain of our two amplifiers (23+37=60 dB) and a few dB of insertion loss. The difference of the transmission and gain curves yields the attenuation to the sample.

\section{Fitting resonator responses}\label{sec:apdx_fittingRoutines}
Because our PCRs are typically under-coupled by a factor of one hundred at small drive powers and low temperatures, the resonant response in $S_{11}$ may become very small compared to the microwave background signal as the probe power is decreased.
Therefore, several hundreds of raw traces must be averaged to overcome the noise floor. To accelerate the data acquisition, we make use of the nonlinear-frequency-sweep procedure presented in Refs.~ \cite{Baity2023, Ganjam2023}, which allows us to decrease the number of frequency points in the sweep by sampling the resonance more efficiently. This method has also been shown to minimize fit bias and the overall fitting error, which for a standard sweep is known to increase with the measurement span $\Delta F$ as $\sigma_{Q_i}\propto \sqrt{\Delta F/\kappa}$. Standard VNA measurements rely on frequency points linearly distributed across the measurement span $\Delta F$. When projecting the resonance data onto the complex plane, this results in a non-uniform point distribution over the circumference of the circle, with a higher density at the off-resonant points of the circle. In such a measurement scheme, most of the data points are therefore not useful to the circle fitting and actually introduce potential bias. Alternatively, one can construct a ``homophasal point distribution" $\{f_n\}_{n\in[-(N-1)/2,\, (N-1)/2]}$ by designing a nonlinear frequency sweep, where the frequency spacing increases quadratically away from the center frequency: 
\begin{equation}
   f_n = f_r + \frac{\Delta F}{2}\frac{\tan{\bigl(\frac{n}{N-1}\Delta\theta\bigr)}}{\tan{(\Delta\theta/2)}}\;\;\textrm{with}\;\Delta\theta = \arctan{\frac{2W}{1-W^2}},
\end{equation}
where $W=\Delta F/\kappa$, $\Delta F$ is the measurement span around $f_r$ and $\kappa$ the total resonance linewidth. Remarkably, to apply this method only an approximate knowledge of $f_r$ and $\kappa$ is required, as the whole scheme depends only on a single parameter, the frequency-span-to-linewidth ratio $W$, which can be estimated beforehand by performing an initial linear sweep across the resonance. This parameter determines the density of points around resonance and we recover the standard linear scheme by taking the limit $W\to 0$. In practice, this method can be implemented using the arbitrary segment mode of the VNA. We set $N=201$ points and aim at $W\approx 5$.

From the circle fit, we extract the resonance frequency $f_r=\omega_r/2\pi$ as well as the total and external quality factors $Q_t$ and $Q_e$. The mean resonance phonon number $\overline{n}$ can then be deduced using
\begin{equation}
    \overline{n} = \frac{2}{\hbar\omega_r^2}\frac{Q_t^2}{Q_e}P_\textrm{in}.
\end{equation}
Importantly, $P_\textrm{in}$ is defined as the power at the sample and not the output power of the VNA, $P_\textrm{VNA}$. Knowing $P_\textrm{VNA}$, $P_\textrm{in}$ is deduced by a careful calibration of the fridge input attenuation using a VTS calibration (see Appendix~\ref{sec:apdx_VTScalib}).

\section{Standard TLS model \& derivation of the resonant and relaxation damping terms}\label{apdx: TLS model}
In this section we will review the standard TLS model and, starting from the harmonic oscillator-TLS interaction Hamiltonian, derive absorption and frequency shift terms resulting from both resonant and relaxation interactions. 

\subsection{TLS interaction Hamiltonian}\label{sec:TLSinteracH}
We start from the well-studied tunneling states (TS) Hamiltonian cast in the non-diagonal, localized representation \cite{Phillips1987}:
\begin{equation}\label{HTS}
    H_{TS} = -\frac{\Delta}{2}\sigma_z + \frac{\Delta_0}{2}\sigma_x,
\end{equation}
which is parameterized by two energy scales: 
the asymmetry energy $\Delta$, setting the energy difference between the lowest energy level in each of the two potential wells defining the TS, and the tunneling energy $\Delta_0$, related to the energy barrier between the two wells. 

Experiments with TLS show dramatic changes in the TLS energy under the application of mechanical strain to the sample \cite{Grabovskij2012}. Strain modifies the asymmetry energy between the two wells, leading to a linear coupling between the TLS and phonons. A perturbation $\delta\xi\ll\xi_0$ in the strain field $\xi=\xi_0+\delta\xi$ induces a change $\delta\Delta\ll\Delta$ in the asymmetry energy. Expanding (\ref{HTS}) to first order in the perturbation $\delta\xi$, we see
\begin{equation}\label{HTSph}
    H_{TS}(\xi) \approx -\frac{1}{2}\left(\Delta(\xi_0) + \evalat*{\frac{\partial \Delta}{\partial \xi}}{\xi=\xi_0}\mathclap{\delta\xi}\;\right)\sigma_z + \frac{\Delta_0}{2}\sigma_x.
\end{equation}
To simplify the notation, we rewrite the static contribution $\Delta(\xi_0)=\Delta$ and introduce the deformation potential $\gamma_z\equiv-(1/2)\partial\Delta/\partial\xi$.
The frequency needed to induce resonant transitions from one level of the TS to the other is typically $\lesssim 20~$GHz for states contributing to thermal properties below 1K \cite{Phillips1987}. The corresponding phonon wavelength is on the order of $100~$nm -- much larger than the spatial extent of the TS. Therefore, when interacting with such a slowly spatially-varying strain field $\delta\xi$, the TS defect can be reasonably treated as point-like and $\gamma_z$ can be understood as the elastic dipole moment of the TS.

We diagonalize (\ref{HTSph}) by introducing the transfer matrix $P$, which yields the interaction Hamiltonian
\begin{align}\label{HTSphdiag}
    H_{TS-ph} &= P^{-1}H_{TLS-ph}P\nonumber\\
    &= -\frac{E}{2}\sigma_z + \left(\frac{\Delta}{E}\sigma_z+\frac{\Delta_0}{E}\sigma_x\right)\gamma_z\;\delta\xi
\end{align}
where P=
$\begin{pmatrix}
\cos(\theta/2) & -\sin(\theta/2)\\
\sin(\theta/2) & \cos(\theta/2)
\end{pmatrix}$ with $\tan(\theta)=\frac{\Delta_0}{\Delta}$, and we define the energy eigenstates $E=\sqrt{\Delta^2+\Delta_0^2}$. Quantizing the strain field $\delta\xi = \xi_{vac}(\hat{a}^\dag+\hat{a})$ by using the phonon creation and annihilation operators $\hat{a}^\dag$ and $\hat{a}$, and explicitly including the phonon energy term $H_{ph}=\hbar\omega_r(\hat{a}^\dag \hat{a}+1/2)$, with $\omega_r$ the phonon frequency, we obtain the full TS-phonon Hamiltonian:
\begin{multline}\label{Hint1}
    H_{TS-ph} = -\frac{E}{2}\sigma_z + \hbar\omega_r\left(\hat{a}^\dag \hat{a} + \frac{1}{2}\right)\\ + \hbar\left(g_z\sigma_z+g_x\sigma_x\right)(\hat{a}^\dag + \hat{a})
\end{multline}
for which we define the longitudinal and transverse coupling strengths:
\begin{align}[left=\empheqlbrace]\label{coupstrength}
    \begin{split}
        g_z &= \frac{\gamma_z}{\hbar}\frac{\Delta}{E}\xi_{vac}\\
        g_x &= \frac{\gamma_z}{\hbar}\frac{\Delta_0}{E}\xi_{vac}
    \end{split}
\end{align}
We have thus recovered the interaction Hamiltonian used by Phillips \cite{Phillips1987} and more recently in Refs. \cite{MacCabe2020, Chen2023}. Note that this is the well-known Rabi Hamiltonian augmented with a longitudinal $\sigma_z$ interaction term \cite{Braak2011}. In this TS model, the ratio of the longitudinal to transverse couplings is given by $\Delta/\Delta_0$. There exists a more general formulation -- known as the \textit{two-level systems} (TLS) model -- in which both couplings are treated as independent quantities that do not depend on the TLS energy. The equivalent TLS interaction Hamiltonian is given by
\begin{equation}
    H_{TLS-ph} = \left(D\sigma_z + M\sigma_x\right)\xi_{vac}(\hat{a}^\dag + \hat{a}),
\end{equation}
where we label $M$ ($D$) the transverse (longitudinal) coupling strength. We will use the terms TS or TLS interchangeably. 

\subsection{Acoustic susceptibility due to a single TLS}\label{sec:acoustSuscep}

We will now derive the finite-frequency response associated with the interaction Hamiltonian in (\ref{Hint1}). From linear response theory, one can compute a generalized acoustic susceptibility $\chi(\omega)$ which contains the response of a TLS bath to some time-varying strain-field. From this susceptibility one can then deduce all relevant quantities such as the TLS-state dependent frequency shift $\delta\omega_r$ experienced by the phonon mode as well as its energy damping rate $\kappa_r$:
\begin{align}[left=\empheqlbrace]
    &\delta\omega_r = -\frac{\xi_\textrm{vac}^2}{\hbar}\Re [\chi(\omega_r)]\label{eq:apdx_acousticChi_fshiftTLS}\\
    &\kappa_r = \frac{2\xi_\textrm{vac}^2}{\hbar}\Im [\chi(\omega_r)],\label{eq:apdx_acousticChi_kappaTLS}
\end{align}
where $\xi_\textrm{vac}=\sqrt{\hbar\omega_r/(2\rho \overline{v}^2 V)}$ is the vacuum strain amplitude ($\overline{v}$ is the average acoustic velocity in the material, $\rho$ is the bulk material mass density, and $V$ is the material volume). 

Trif \textit{et al.} derived a useful formula \cite{Trif2018, Dassonneville2014} for computing the dynamical susceptibility $\chi(\omega)$ of some Hamiltonian $H(t)$ in the presence DC and AC perturbations; we apply this formalism to the case of a strain field perturbation $\xi(t)$ acting on a single TLS. We begin by decomposing $H(t)$ into a static and a time-dependent contribution (to leading order in the AC strain field $\delta\xi(t)$),
\begin{align}[left=\empheqlbrace]\label{Hsusc}
    H(t) &= H_0 + V(t) \\
    V(t) &= -\hat{p}\,\delta\xi(t) = \frac{\partial H_0}{\partial\xi}\delta\xi(t)\nonumber,
\end{align}
where $H_0 = H_{TLS}+H_{ph} $ is the Hamiltonian in the absence of the AC strain field, and $\hat{p}$ is the static elastic dipole of the TLS in absence of the drive. Comparing this expression with (\ref{HTSphdiag}), we find
\begin{equation}\label{dipOp}
    \hat{p} = -\gamma_z\left(\frac{\Delta}{E}\sigma_z+\frac{\Delta_0}{E}\sigma_x\right).
\end{equation}
The dynamics of the TLS are characterized by the evolution of its density operator $\rho(t)$ under $H(t)$. We must include a term in $\rho(t)$ to account for the relaxation of the TLS to its environment, a process mostly mediated by thermal phonons \cite{Trivedi1988}:
\begin{equation}\label{rho}
    \frac{\partial\rho}{\partial t} + \frac{i}{\hbar}[H(t), \rho(t)]=-\hat{\Gamma}[\rho(t)-\rho_{eq}(t)],
\end{equation}
where $\rho(t)$ is then to be understood as the reduced density operator of the TLS after tracing over the environment. $\hat{\Gamma}$ is the TLS relaxation tensor (containing both diagonal and off-diagonal relaxations), and $\rho_{eq}(t)=\exp{(H(t)/k_B T)}/Z$ is the quasi-equilibrium density operator, with $Z=\textrm{Tr}[\exp{(H(t)/k_B T)}]$ the grand partition function. For now, the term describing relaxation to equilibrium can be viewed as a phenomenological addition to account for dissipation. We will later derive it using the Lindblad master equation and remark on the approximations behind its derivation. Using (\ref{Hsusc})-(\ref{rho}), Trif \textit{et al.} derived a general recipe to compute the susceptibility $\chi(\omega)=\delta\langle\hat{p}\rangle/\delta\xi(t)$, where $\langle\hat{p}\rangle=\text{Tr}[\hat{p}(t)\rho(t)]$:
\begin{align}\label{chi}
    &\chi(\omega)=\sum_n \chi^{(n)}(\omega),\\
    &\chi^{(n)}(\omega) = \underbrace{-f_n\frac{\partial^2 \epsilon_n}{\partial\xi^2}}_{\chi_C^{(n)}} \; \underbrace{-\;\frac{i\Gamma_{nn}}{\omega + i\Gamma_{nn}}\left(\frac{\partial\epsilon_n}{\partial\xi}\right)^2\frac{\partial f_n}{\partial \epsilon_n}}_{\chi_D^{(n)}}\nonumber\\
    &\;\underbrace{-\hbar\omega\sum_{m\neq n}\frac{|\langle m|\frac{\partial H_0}{\partial\xi}|n\rangle|^2}{\epsilon_n-\epsilon_m}\frac{f_n-f_m}{\epsilon_n-\epsilon_m-\hbar\omega-i\hbar\Gamma_{nm}}}_{\chi_{ND}^{(n)}},\nonumber
\end{align}
for which $|n\rangle$ and $\epsilon_n$ are eigenvectors and eigenstates of the bare Hamiltonian $H_0$ in the absence of the driving, such that $H_0|n\rangle=\epsilon_n|n\rangle$ and $\rho_0|n\rangle=f_n|n\rangle$, and the sum is carried out over all the states \cite{Trif2018}.
The susceptibility can thus be decomposed into three
parts: the \textit{curvature} $\chi_C$, \textit{diagonal} $\chi_D$, and \textit{non-diagonal} $\chi_{ND}$ (or Kubo) contributions, respectively. We will show that $\chi_{ND}$ is responsible for the `resonant' TLS damping and frequency shift (arising from the $\sigma_x$ term in the interaction Hamiltonian), while $\chi_D$ corresponds to the `relaxation' TLS damping. 

The curvature contribution can be related to the adiabatic susceptibility $\chi_A=\partial P/\partial\xi$ with $P=\sum_n f_n\langle n|\hat{p}|n\rangle = -\sum_n f_n\partial\epsilon_n/\partial\xi$:
\begin{equation}
    \chi_A = \chi_C-\sum_n \frac{\partial f_n}{\partial\epsilon_n}\left(\frac{\partial\epsilon_n}{\partial\xi}\right)^2
\end{equation}
and so,
\begin{equation}
    \chi_C+\chi_D = \chi_A + \sum_n \frac{\partial f_n}{\partial\epsilon_n}\left(\frac{\partial\epsilon_n}{\partial\xi}\right)^2\frac{\omega}{\omega+i\Gamma_{nn}}.
\end{equation}
This validates the definition of $\chi_A$ as the adiabatic susceptibility; when taking the adiabatic limit $\omega\to 0$, $\chi_{ND}\to 0$ and $\chi\to\chi_A=\chi_C+\chi_D$. For a closed system in which the populations are fixed, $\partial f_n/\partial\epsilon_n=0$ and $\chi_A=\chi_C\propto \partial^2\epsilon_n/\partial\xi^2$\nolinebreak: the total susceptibility is given by the curvature of the energy levels. 

We are now ready to compute $\chi(\omega)$ for the TLS-phonon Hamiltonian. 
We label the two TLS states as `1' and `2, with energies $\epsilon_1=-E/2$ and $\epsilon_2=E/2$. The equilibrium TLS populations are then given by
\begin{align}\label{f1}
    &f_1 = \frac{e^{-\epsilon_1/k_B T}}{e^{-\epsilon_1/k_B T}+e^{-\epsilon_2/k_B T}} = \frac{1}{1+e^{-E/k_B T}}\nonumber\\
    &f_2 = 1-f_1,
\end{align}
from which we deduce the equilibrium population difference $\langle \sigma_z\rangle_{eq} = \textrm{Tr}[\sigma_z\rho_{eq}]$:
\begin{align}\label{popdiff}
    \langle \sigma_z\rangle_{eq} &= \frac{1-e^{-E/k_B T}}{1+e^{E/k_B T}} = \tanh{\left(\frac{E}{2 k_B T}\right)}.
\end{align}
We also derive a few useful identities which will help simplifying expressions. Starting from $E=\sqrt{\Delta^2+\Delta_0^2}$ and taking successive derivatives with respect to strain $\xi$, one obtains:
\begin{align}\label{Ederivatives}
    \frac{\partial E}{\partial\xi} &= -2\gamma_z\frac{\partial E}{\partial\Delta} = -2\gamma_z\frac{\Delta}{E}\\
    \frac{\partial^2 E}{\partial\xi^2} &= -2\frac{\partial\gamma_z}{\partial\xi}\frac{\Delta}{E}+\frac{4\gamma_z^2}{E}-2\gamma_z\Delta\frac{2\gamma_z\Delta}{E^3}\nonumber\\
    &= \frac{\Delta}{E}\frac{\partial^2\Delta}{\partial\xi^2}+4\gamma_z^2\frac{\Delta_0^2}{E^3}
\end{align}
From the second line, we see \footnote{This identity is a particular case of a more general type of expressions known as a \textit{sum-rule}. It relates the diagonal matrix elements of the second derivative of the Hamiltonian with respect to an external perturbation (strain, flux, etc) to the energy curvature of the energy levels and the matrix elements of the first derivative, and can be derived in this general form using the Hellmann-Feynman theorem \cite{Metzger2022}
\begin{equation*}
    \langle i| H''|i\rangle = E_i''+2\sum_{j\neq i}\frac{|\langle j|H'|i\rangle|^2}{E_j-E_i}
\end{equation*}
}:
\begin{equation}\label{sumrule}
    \frac{\Delta}{E}\frac{\partial^2\Delta}{\partial\xi^2}=\frac{\partial^2E}{\partial\xi^2}-4\frac{\hbar^2 g_x^2}{E \xi_{vac}^2}
\end{equation}

We now work out each term in (\ref{chi}), beginning with the non-diagonal term $\chi_{ND}$. We introduce $f_{nm}=f_n-f_m$ and $\epsilon_{nm}=\epsilon_n-\epsilon_m$, and neglect for now the transverse relaxation $\Gamma_2\equiv\Gamma_{nm} =\Gamma_{mn}\approx0$. We rewrite $\chi_{ND}$ into a form which will allow us to make contact with the frequency shift,
\begin{multline}\label{chiND}
    \chi_{ND} = \frac{1}{2}\sum_{m\neq n}|\langle m|\frac{\partial H_0}{\partial\xi}|n\rangle|^2 f_{nm} \\
    \times \Biggl[\frac{2}{\epsilon_{nm}}-\frac{1}{\epsilon_{nm}-\hbar\omega}+\frac{1}{\epsilon_{nm}+\hbar\omega}\Biggr].
\end{multline}
    
From (\ref{dipOp}),(\ref{coupstrength}), and (\ref{Ederivatives}), we then compute the matrix elements between the TLS states:
\begin{align}[left=\empheqlbrace]
    \langle i|\frac{\partial H_0}{\partial\xi}|i\rangle &= \langle i|\gamma_z\frac{\Delta}{E}\sigma_z|i\rangle = \pm\frac{\hbar g_z}{\xi_{vac}}\label{derHz}\\
    \langle2|\frac{\partial H_0}{\partial\xi}|1\rangle &=\langle2|\gamma_z\frac{\Delta_0}{E}\sigma_x|1\rangle = \frac{\hbar g_x}{\xi_{vac}}\label{derHx}
\end{align}

Plugging (\ref{derHx}) into (\ref{chiND}), we obtain:
\begin{multline} \label{chiND2}
    \chi_{ND} = -\Biggl(\frac{\hbar g_x}{\xi_{vac}}\Biggr)^2 \tanh{\left(\frac{E}{2 k_B T}\right)} \\
    \times \Biggl[\frac{2}{E}-\frac{1}{E-\hbar\omega}-\frac{1}{E+\hbar\omega} \Biggr] \\
\end{multline}
This term describes the finite-frequency response due to virtual TLS transitions mediated by the absorption ($\hbar\omega=E$) and emission ($\hbar\omega=-E$) of phonons. An alternative method to derive it is to approximately diagonalize the Hamiltonian \ref{Hint1} by applying a Schrieffer-Wolff unitary transformation $\mathcal{U}=\exp{[(\gamma(\hat{a}^\dag\sigma_- -\hat{a}\sigma_+)-\overline{\gamma}(\hat{a}^\dag\sigma_+ - a\sigma_-)]}$ with $\gamma=g_x/(E-\hbar\omega_r)$ and $\overline{\gamma}=g_x/(E+\hbar\omega_r)$ and to expand up to second order in $g_x$ using the Baker-Campbell-Hausdorff lemma \cite{Zueco2009, Billangeon2015}. This expression is valid in the full dispersive regime $g_x\ll|E-\hbar\omega_r|$ beyond the rotating-wave approximation. 

The curvature term is also easily obtained:
\begin{equation}\label{chiCurv}
    \chi_C = \frac{1}{2}\tanh{\left(\frac{E}{2 k_B T}\right)}\frac{\partial^2 E}{\partial \xi^2}.
\end{equation}

To find the diagonal susceptibility, we define $\Gamma_1\equiv\Gamma_{11}=\Gamma_{22}$, and see
\begin{align}\label{chiDiag}
    \begin{split}
        \chi_D&= \frac{1}{1-i\omega\Gamma_1^{-1}}\frac{1}{2}\Biggl(\frac{\partial E}{\partial\xi}\Biggr)^2\frac{\partial f_{12}}{\partial E} \\
        &=\frac{\gamma_z^2}{k_B T}\frac{\Delta^2}{E^2}\frac{\textrm{sech}^2\frac{E}{2 k_B T}}{1-i\omega\Gamma_1^{-1}}.
    \end{split}
\end{align}
This expression coincides exactly with the longitudinal  (relaxation) susceptibility derived by Phillips (see (4.32),(4.33) in Ref. \cite{Phillips1987}) and later re-derived by MacCabe \textit{et al.} (see (S-81) in Suppl. Ref.~ \cite{MacCabe2020}).

Finally collecting the susceptibilities into (\ref{eq:apdx_acousticChi_fshiftTLS}), we arrive at the frequency shift $\delta\omega_r$ of the phonon mode due to its coupling to a single TLS:
\begin{multline}\label{fshift2}
    \hbar\delta\omega_r = -\frac{1}{2}\frac{\partial^2E}{\partial\xi^2}\tanh{\left(\frac{E}{2 k_B T}\right)}\;\xi_{vac}^2\; \\
    - \;\frac{\hbar^2 g_z^2}{k_B T}\frac{\textrm{sech}\frac{E}{2 k_B T}^2}{1+(\omega_r\Gamma_1^{-1})^2} +\hbar^2 g_x^2 \tanh{\left(\frac{E}{2 k_B T}\right)} \\
    \times \Biggl(\frac{2}{E}-\frac{1}{E-\hbar\omega_r}-\frac{1}{E+\hbar\omega_r}\Biggr).
\end{multline}
In this expression, all three terms are proportional to $\xi_{vac}^2$, so this frequency shift effectively corresponds to a second-order expansion in the perturbation $\delta\xi$. Using the identity (\ref{sumrule}), we rewrite (\ref{fshift2}),
\begin{multline}\label{fshift3}
    \hbar\delta\omega_r = -\frac{\xi_{vac}^2}{2}\frac{\Delta}{E}\frac{\partial^2\Delta}{\partial\xi^2}\tanh{\left(\frac{E}{2 k_B T}\right)}\; \\
    - \;\frac{\hbar^2g_z^2}{k_B T}\frac{\textrm{sech}^2\frac{E}{2 k_B T}}{1+(\omega_r\Gamma_1^{-1})^2}
    - \;\hbar^2 g_x^2 \tanh{\left(\frac{E}{2 k_B T}\right)} \\
    \times \Re\Biggl\{\frac{1}{E-\hbar\omega_r-i\hbar\Gamma_2}+\frac{1}{E+\hbar\omega_r+i\hbar\Gamma_2}\Biggr\},
\end{multline}
where we have also restored the transverse relaxation terms $\Gamma_2$ in the non-diagonal term. We recover the same expression derived by MacCabe \textit{et al.} \cite{MacCabe2020}, plus an extra contribution: the first term, corresponding to the \textit{adiabatic} shift. It appears in our derivation because we performed the expansion up to second order in $\delta\xi$, but can be considered negligible in practice. The second term corresponds to the \textit{TLS relaxation} contribution which arises from longitudinal coupling. The last term is the usual \textit{resonant} (dispersive) contribution, due to the transverse interaction with the TLS. 

\subsection{Resonant TLS damping \& frequency shift}\label{sec:resonantTLS}
In the previous section, we derived the susceptibility for a single TLS interacting with the phonon mode. In reality, acoustic resonators are coupled to a bath of TLS with a broad energy spectrum which necessitates summing over all TLS contributions in (\ref{chiND2})-(\ref{chiDiag}).
The non-diagonal susceptibility $\chi_{ND}$ then reads:
\begin{multline}\label{eq:apdx_acousticChi_chiNDSum}
    \hspace*{-0.5cm}\chi_{ND}(\omega) = \frac{\hbar^2 P V_h}{\xi_\textrm{vac}^2}
    \int_{0}^{\omega_\textrm{max}}\mathrm{d}\omega_\textrm{TLS}\,g_x^2\tanh{\frac{\hbar\omega_\textrm{TLS}}{2 k_B T}}\times\\
    \biggl(\frac{1}{\omega_\textrm{TLS}-(\omega+i\Gamma_2^\textrm{TLS})}
    +\frac{1}{\omega_\textrm{TLS}+\omega+i\Gamma_2^\textrm{TLS}}\biggr),
\end{multline}
for which we introduce the cutoff frequency $\omega_\textrm{max}$ to describe the maximum transition frequency of the TLS ensemble. We also assume $\partial^2\Delta/\partial\xi^2\approx 0$, which causes the adiabatic susceptibility to cancel the first term in (\ref{chiND2}). This is the same expression as the complex susceptibility that was computed by Hunklinger \& Arnold \cite{Hunklinger1976}. In the continuum limit, the discrete sum over TLS can be replaced by an integral over the TLS local density of states $P$, $\sum_\textrm{TLS}\longleftrightarrow\int d V\int d E_\textrm{TLS} P$, which reduces to $\hbar P V_h\int d\omega_\textrm{TLS}$ in the case that the DOS is flat in energy and TLS are uniformly distributed in a volume $V_h$ of the host material.

Assuming $\hbar\omega_\textrm{max}\gg k_B T$, the integral can be conveniently evaluated using an integral representation of the complex digamma function $\Psi$ \cite{AbramowitzDigamma1964}:
\begin{multline}\label{eq:apdx_acousticChi_chiDigamma}
    \chi_{ND}(\omega_r) \approx -2 P V_h \overline{M}^2\biggl( \Psi\Bigl(\frac{1}{2}+\frac{\hbar\Gamma_2^\mathrm{TLS}}{2\pi k_B T}\\
    +\frac{\hbar\omega_r}{2\pi i k_B T}\Bigr)
    -\mathrm{ln}\Bigl(\frac{\hbar\omega_\mathrm{max}}{2\pi k_B T}\Bigr)\biggr),
\end{multline}
for which we introduce the averaged (over TLS orientation and acoustic polarization) transverse coupling potential $\overline{M}$, which is related to the coupling rate by $\overline{g_x}=(\overline{M}/\hbar)\xi_\textrm{vac}$. Doing this, we effectively assume a TLS (rather than TS) model, since the energy dependence of the coupling constants is neglected.

In practice, $\hbar\Gamma_2^\textrm{TLS}/2\pi k_B T\ll 1/2$ and can therefore be neglected. The loss $\kappa_r$ is then directly obtained by injecting (\ref{eq:apdx_acousticChi_chiDigamma}) into (\ref{eq:apdx_acousticChi_kappaTLS}) and employing the remarkable identity $\Im \Psi(1/2+x/(2\pi i))=-(\pi/2)\tanh{x/2}$:
\begin{equation}
    \kappa_r = \frac{2}{\hbar}\cdot\frac{\hbar\omega_r}{2\rho\overline{v}^2 V}\cdot2 P V_h\overline{M}^2\cdot\frac{\pi}{2}\tanh{\frac{\hbar\omega_r}{2 k_B T}},
\end{equation}
from which we deduce the expression for the resonator inverse quality factor due to resonant TLS coupling:
\begin{equation}\label{eq:apdx_acousticChi_QTLS}
    Q_\textrm{TLS}^{-1}=\frac{\kappa_r}{\omega_r}= F\delta^0\tanh{\frac{\hbar\omega_r}{2 k_B T}}\quad\textrm{with}\;\delta^0=\frac{\pi P \overline{M}^2}{\rho\overline{v}^2},
\end{equation}
where $\delta^0$ is the intrinsic TLS loss tangent at zero temperature and $F=V_h/V$ is the filling fraction of TLS in the host material.

 Using the relation between the deformation potential $\overline{M}$ and the transverse coupling strength $\overline{g_x}$ and that $\xi_\textrm{vac}=\sqrt{\hbar\omega_r/(2\rho\overline{v}^2 V)}$, we can re-express the effective loss tangent as $F\delta_0 = 2\pi V_h P\hbar \overline{g_x}^2 /\omega_r$. Introducing the TLS frequency DOS $\rho_\omega$, which we can relate to the TLS local energy DOS $P$ from $\int\rho_\omega(\omega)d\omega = \int dV\int P(E)dE=\hbar V_h \int P(E) d\omega$, we obtain
 \begin{equation}
     \kappa_r(T=0) =\omega_r F\delta_0 =2\pi \rho_\omega \overline{g_x}^2.
 \end{equation}
 This is the Fermi golden rule result expected for coupling to a dense TLS ensemble. For coupling to a dilute TLS ensemble, we would expect instead the Purcell limit $\kappa_r \propto \Gamma_{1,\textrm{TLS}} (\overline{g_x}/\Delta)^2$ with $\Delta$ the detuning between the resonator and the nearest TLS, as discussed in Refs.~ \cite{Odeh2023,Spiecker2023b}.

 Similarly, using (\ref{eq:apdx_acousticChi_fshiftTLS}) and (\ref{eq:apdx_acousticChi_chiDigamma}), we obtain the expression for the resonator's relative frequency shift due to its dispersive interaction with a continuum of TLS:
 \begin{equation}\label{eq:apdx_acousticChi_domegar}
     \frac{\delta\omega_r}{\omega_r}=\frac{F\delta^0}{\pi}\biggl(\Re \Bigl\{\Psi\Bigl(\frac{1}{2}+\frac{\hbar\omega_r}{2\pi i k_B T}\Bigr)\Bigr\}-\textrm{ln}\Bigl(\frac{\hbar\omega_\textrm{max}}{2\pi k_B T}\Bigr)\biggr)
 \end{equation}
Using the asymptotic expansion of the digamma function $\Psi(z)\sim \ln{(z)}$ as $z\rightarrow\infty$ (equivalent to the limit $T\rightarrow 0$) we can get rid of the TLS cutoff frequency $\omega_\textrm{max}$ and obtain a simpler expression for the resonator shift by expressing it \textit{with respect to the bare resonant frequency at $T=0$}, as was done in Refs.~ \cite{Phillips1987, Hunklinger1976}:
\begin{equation}
    \frac{\Delta\omega_r}{\omega_r}\equiv\frac{\omega_r(T)-\omega_r(0)}{\omega_r(\infty)}= \frac{\delta\omega_r(T)}{\omega_r}-\frac{\delta\omega_r(0)}{\omega_r},
\end{equation}
which finally yields (\ref{eq:resTLS_freqShift}) from the main text:
\begin{equation}\label{dfTLS}
    \frac{\Delta\omega_r}{\omega_r}=\frac{F\delta^0}{\pi}\Bigl[\Re \Bigl\{\Psi\Bigl(\frac{1}{2}+\frac{\hbar\omega_r}{2\pi i k_B T}\Bigr)\Bigr\}-\textrm{ln}\Bigl(\frac{\hbar\omega_r}{2\pi k_B T}\Bigr)\Bigr]
\end{equation}

To stress that this relative frequency shift is expressed with respect to the frequency at zero temperature, $\omega_r(0)$, we use a different notation $\Delta\omega_r$ instead of $\delta\omega_r=\omega_r(T)-\omega_r$, where the bare resonator frequency, \textit{i.e.} its frequency in absence of coupling to the TLS, corresponds to the limit $\omega_r\equiv\omega_r(T\rightarrow\infty)$. Indeed, at infinite temperature, all the TLS are saturated: their two levels are populated equally, $\langle\sigma_z\rangle=\tanh{[\hbar\omega_\textrm{TLS}/2k_B T]}\underset{T\to\infty}{\rightarrow}0$, and so the susceptibility in (\ref{eq:apdx_acousticChi_chiNDSum}) vanishes: their dispersive shift on the resonator, which are opposite to each other, cancel perfectly, thus giving no net shift. Approximating $\Psi(z)\underset{z\to\infty}{\sim} \ln{(z)}$, one can directly verify from (\ref{dfTLS}) that $\lim_{T \rightarrow 0} \Delta \omega_r = 0$. The physical limit $\lim_{T \rightarrow \infty} \Delta \omega_r = (\omega_r-\omega_r(0))/\omega_r$ is however not recovered by (\ref{dfTLS}), which instead predicts a logarithmic divergence with temperature. This clarifies the limit of validity of (\ref{dfTLS}) to be $\omega_\textrm{TLS}\ll\omega_\textrm{max}$. In practice, $\omega_\textrm{max}\gg\omega_r$ and (\ref{dfTLS}) remains generally valid.

\subsection{Power dependence of the resonant damping term \& effect of TLS-TLS interactions}

The expression that we derived for the resonant TLS damping is valid in the weak resonator field limit where the TLS population is only thermal and is not driven by the resonator field. At high phonon number, however, we must take into account the TLS coherence by solving Bloch equations. When probing the acoustic resonator at $\omega\approx\omega_r$, the generated strain field $2\xi_0\sin{\omega t}$ will couple to and saturate near-resonant TLSs through the elastic dipole operator. The population of a TLS excited state at saturation is given by the steady-state solution to Bloch equations \cite{Palacios2010}:
\begin{equation}\label{pesat}
    p_e^\textrm{sat} = \frac{1}{2}-\Bigl(\frac{1}{2}-p_e^\textrm{th}\Bigr) \frac{1+\Bigl(\frac{\delta\omega}{\Gamma_2}\Bigr)^2}{1+\Bigl(\frac{\delta\omega}{\Gamma_2}\Bigr)^2+\Bigl(\frac{\Omega_R}{\sqrt{\Gamma_1\Gamma_2}}\Bigr)^2},
\end{equation}
where $\Gamma_1=1/T_1$ and $\Gamma_2=1/T_2$ are the TLS transverse and longitudinal relaxation rates, $p_e^\textrm{th}$ is the thermal population of the TLS excited state, $\delta\omega=\omega_\textrm{TLS}-\omega$ is the detuning between the TLS and the drive frequency and $\Omega_R=2 |p_{12}|\xi_0/\hbar=2 \overline{g_x}$ is the TLS Rabi frequency, for which $|p_{12}|=\gamma_z\Delta_0/E$ is the TLS elastic dipole. This coincides with what we defined as the transverse coupling potential $\overline{M}$ in Section~\ref{sec:resonantTLS}.

From (\ref{pesat}), we can now express the steady-state population difference $\Delta p$ between the TLS ground and excited states in presence of a driving field on the resonator, assuming $p_e+p_g=1$:
\begin{equation}
    \Delta p = \Delta p^\textrm{th}\frac{1+\Bigl(\frac{\delta\omega}{\Gamma_2}\Bigr)^2}{1+\Bigl(\frac{\delta\omega}{\Gamma_2}\Bigr)^2+\Bigl(\frac{\Omega_R}{\sqrt{\Gamma_1\Gamma_2}}\Bigr)^2},
\end{equation}
where $\Delta p^\textrm{th}=2p_e^\textrm{th}-1=\langle\sigma_z\rangle_\textrm{eq}=\tanh{[\hbar\omega_\textrm{TLS}/2k_B T]}$ is the thermal TLS population difference. In the strong field limit, one should then replace $\Delta p^\textrm{th}$ in  (\ref{eq:apdx_acousticChi_chiNDSum}) by $\Delta p$. This expression coincides with (4.20) from Phillips \cite{Phillips1987} and the following discussion is greatly inspired from his derivation. We recast it as a Lorentzian function of the variable $\delta\omega$:
\begin{align}
    \Delta p &= \Delta p^\textrm{th}\frac{A}{1+\Bigl(\frac{\delta\omega}{\Gamma_2'}\Bigr)^2}\\
    &\textrm{with}\;\left\{\begin{aligned}
    A &= \frac{1+(\delta\omega/\Gamma_2)^2}{1+(\Omega_R/\sqrt{\Gamma_1\Gamma_2})^2}\\
    \Gamma_2' &= \Gamma_2\sqrt{1+(\Omega_R/\sqrt{\Gamma_1\Gamma_2})^2}\\
    \end{aligned}\nonumber\right.
\end{align}
This shows that the response on resonance ($\delta\omega=0$) is reduced with large fields by a factor $1+\Omega_R^2 T_1 T_2$ and that the
response curve is broadened, defining an effective relaxation time $T_2'$ where
\begin{equation}\label{T2prime}
    \frac{1}{T_2'^2}=\frac{1}{T_2^2}+\Omega_R^2\frac{T_1}{T_2}
\end{equation}
From (\ref{T2prime}), we rewrite the response reduction at resonance as $1+\Omega_R^2 T_1 T_2=(T_2/T_2')^2$. The TLS contributing to the acoustic loss are the ones within a linewidth of the resonator frequency, $|\omega-\omega_r|\lesssim \kappa_r$. The number of these states, of order $P(\hbar/T_2')$, is increased at large field by a factor $T_2/T_2'$ (because of the response curve broadening). However, the contribution of each is also reduced by the factor $(T_2/T_2')^2$ that we just derived, so that altogether, the effect of the resonator field is to reduce the acoustic attenuation by a factor $T_2/T_2'=\sqrt{1+\Omega_R^2 T_1 T_2}$.
Therefore, the thermal population difference $\Delta p^\textrm{th}$ in (\ref{eq:apdx_acousticChi_QTLS}) should be reduced by this factor to account for the power-dependent TLS saturation from the resonator fiel. We find
\begin{align}\label{QTLSint}
    Q_\textrm{TLS}^{-1}&=\frac{\kappa_r}{\omega_r}= \frac{F\delta^0}{\sqrt{1+\frac{I}{I_c}}}\tanh{\frac{\hbar\omega_r}{2 k_B T}}\\
    &\textrm{with}\;\frac{I}{I_c}=\Omega_R^2 T_1 T_2 = \Bigl(\frac{2\xi_0\overline{M}}{\hbar}\Bigr)^2 T_1 T_2\nonumber.
\end{align}
For a strain field $2\xi_0\sin{\omega t}$, the associated acoustic intensity is given by $I=\overline{v}\mathscr E$ with the average energy density $\mathscr E=\langle\sigma\xi\rangle=Y\langle\xi^2\rangle=2\rho\overline{v}^2\xi_0^2$, where we used the Young modulus $Y = \overline{v}^2\rho$, so that $I=2\rho\xi_0^2\overline{v}^3$. The critical intensity is then $I_c=\hbar^2\rho\overline{v}^3/(2\overline{M}^2 T_1 T_2)$. This intensity ratio can be rewritten as a mean phonon number ratio $\overline{n}/n_c$, by equating the time-averaged classical energy to the energy carried by phonons: $\overline{n}\hbar\omega_r=2\rho\overline{v}^2\xi_0^2V$ with $V$ the resonator mode volume, so that: 
\begin{align}\label{QTLSnbar}
    &Q_\textrm{TLS}^{-1}= \frac{F\delta^0}{\sqrt{1+\frac{\overline{n}}{n_c}}}\tanh{\frac{\hbar\omega_r}{2 k_B T}}\\
    &\textrm{with}\;\;\overline{n}=\frac{2\rho\overline{v}^2\xi_0^2V}{\hbar\omega_r}\;,\;n_c=\frac{\hbar\rho\overline{v}^2V}{2\omega_r\overline{M}^2 T_1 T_2}=\frac{1}{(2\overline{g_x})^2 T_1 T_2}\nonumber
\end{align}
This is the usual expression for the resonant TLS damping found in the TLS literature \cite{Hunklinger1976, Wollack2021}. At higher field, when the TLS Rabi frequency is larger than the TLS decoherence rate $\Gamma_2^\textrm{TLS}$, a nonlinear absorption regime is found and the square-root dependence on the acoustic intensity of the loss tangent is no longer valid \cite{Burin2018, Gorgichuk2023}.

We emphasize that the main results of this section, (\ref{dfTLS}) and (\ref{eq:apdx_acousticChi_QTLS}) for the frequency shift and the acoustic loss due coupling to the TLS bath, were derived under the simplifying assumptions that 1) the TLS density of states (DOS) $P$ is uniform in space and constant with energy, 2) that all TLSs couple to the acoustic mode via the same transverse coupling potential $\overline{M}$ and 3) that TLS do not interact with each other. These are obviously crude assumptions, although they facilitate analytical results which accurately describe many experimental data. 

There has been evidence in recent literature to suggest that the TLS DOS is energy dependent \cite{Burnett2014}, $\rho(E)\propto E^\mu$ with $\mu\sim 0.3$, a phenomenon which may arise from TLSs interacting with each other via dipole-dipole couplings. A generalized tunneling model incorporating such TLS-TLS interactions has been developed by Ioffe \& Faoro in Ref. \cite{Faoro2015}. An important result of this model is that the frequency shift is completely insensitive to the presence of fluctuators coupled to resonant TLS, so that (\ref{dfTLS}) should still describe experimental data well, even in the presence of TLS-TLS interactions. However, the power dependence of the loss in a high quality factor resonator is expected to depart from the square-root dependence predicted by (\ref{QTLSnbar}) and to be weaker. Indeed, resonant TLSs may acquire a finite linewidth $\Gamma_2\propto T^{1+\mu}$ due to their interaction with surrounding thermally excited TLS, resulting in a slower absorption of the resonator energy. $\Gamma_2$ is shown to dominate over the decoherence rate $\Gamma_2^{ph}$ due to phonons, however the relaxation due to the interaction with thermally activated TLSs remains negligible and the TLS energy decay rate $\Gamma_1$ is still dominated by the phonon contribution $(\Gamma_1^{ph})^{-1}\propto \tanh{(\hbar\omega_r/(2 k_B T))}$. Incorporating the $\Gamma_1$ dependency on temperature in (\ref{QTLSnbar}), one obtains (\ref{eq:resTLS_Q}) from the main text. 

An additional replacement was performed, $\overline{n}\rightarrow\overline{n}^\beta$, to account for the non-uniform field distribution within the resonator: TLS at different locations may experience a different strain field and saturate differently as the number of phonons increases. Although $\beta$ is a phenomenological addition to the model, this parametrization has proved to fit well experimental data \cite{Wang2009} and the resulting generalized expression for the power-dependent quality factor, (\ref{eq:resTLS_Q}), was used successfully by Crowley \textit{et al.} to describe TLS losses in Tantalum superconducting circuits \cite{Crowley2023}.

\subsection{Relaxation TLS damping \& frequency shift}\label{sec:relaxationTLS}

The diagonal susceptibility $\chi_{D}$ summed over all TLS is given by
\begin{equation}\label{chiDsum}
    \chi_D(\omega) = \frac{\hbar PV}{\xi_\textrm{vac}^2}\int_{0}^{\omega_\textrm{max}}\mathrm{d}\omega_\textrm{TLS}\,\frac{\hbar^2 \overline{g_z}^2}{k_B T}\frac{\textrm{sech}^2\frac{\hbar\omega_\textrm{TLS}}{2 k_B T}}{1-i\omega \Gamma_1^{-1}}.
\end{equation}
As we did for the transverse coupling, we now introduce the averaged longitudinal coupling potential $\overline{D}$, such that $\overline{g_z}=(\overline{D}/\hbar)\xi_\textrm{vac}$. We assume that the main TLS relaxation mechanism is the one-phonon process, $\Gamma_1\approx\Gamma_1^{ph}$, an approximation which, as we motivated in the previous section, should remain valid even in presence of TLS-TLS interactions. The TLS energy decay rate $\Gamma_1$ is given by $\Gamma_1 = \omega_{g\to e}+\omega_{e\to g} = \omega_{g\to e}(1+\exp{(E/(k_B T))})$, where we assume thermal equilibrium to relate the transition probability $\omega_{e\to g}$ to $\omega_{g\to e}$. This transition probability can be computed for a weak strain field by means of Fermi's golden rule:
\begin{equation}
    \omega_{g\to e} = \sum_\alpha \frac{2\pi}{\hbar}|\langle g| H_{TLS-ph} |e\rangle|^2 \frac{\rho_d(\omega_\textrm{TLS})}{\hbar} N(\omega_\textrm{TLS}),
\end{equation}
where $\rho_d$ is the Debye density of states for a $d$-dimensional phonon bath, $N(\omega_\textrm{TLS})=1/(e^{\hbar\omega_\textrm{TLS}/(k_BT)}-1)$ is the phonon distribution at the TLS frequency, and the sum is over the phonon polarization $\alpha$. Averaging quantities over orientations, one can rewrite the transition matrix element $|\langle g| H_{TLS-ph} |e\rangle|\approx \hbar\overline{g_x}=\overline{M}\xi_\textrm{vac}$ and compute the TLS energy decay rate
\begin{align}
    \Gamma_1(\omega_\textrm{TLS}) &= \frac{2^{1-d}\pi^{1-d/2}}{\Gamma(d/2)}\frac{\overline{M}^2\omega_\textrm{TLS}^d}{\hbar\rho\overline{v}^{d+2}S_{3-d}}\coth{\Bigl(\frac{\hbar\omega_\textrm{TLS}}{2k_B T}\Bigr)},
\end{align}
where $S_{3-d}$ is the $(3-d)$ dimensional cross section ($S_2=w^2,\,S_1=t,\,S_0=1$ with $w$ and $t$ the width and thickness of the host material) and we used the following formula for the $d$-dimensional Debye density of states, $\rho_d(\omega)=2V_d/(\Gamma(d/2)(2\sqrt{\pi})^d)(\omega^{d-1}/\overline{v}^d)$ from Ref.~ \cite{Valladares1975} with $V_d$ the $d$-dimensional volume. Having an expression for $\Gamma_1$, we can now compute the decay rate $\kappa_r$ via (\ref{chiDsum}) and (\ref{eq:apdx_acousticChi_kappaTLS}). This is typically considered in two limits, $\omega_r\Gamma_1^{-1}\gg 1$ and $\omega_r\Gamma_1^{-1}\ll 1$; we continue with the assumption of the former for $\omega_r\approx 500$ MHz resonators. We find the decay rate to be
\begin{align}
    & \kappa_r = 2 P V\int_{0}^{\omega_\textrm{max}}\mathrm{d}\omega_\textrm{TLS}\,\frac{\hbar\overline{g_z}^2}{\omega_r}\frac{\hbar\Gamma_1(\omega_\textrm{TLS})}{k_B T}\textrm{sech}^2\frac{\hbar\omega_\textrm{TLS}}{2 k_B T}\nonumber\\
    &=\frac{A_d \overline{D}^2\overline{M}^2 P}{\hbar\rho^2 \overline{v}^{d+4}S_{3-d}}\Bigl(\frac{k_B T}{\hbar}\Bigr)^d\int_{0}^{\quad\mathclap{x_\textrm{max}}}\mathrm{d}x\,\textrm{csch}(x)x^d,
\end{align}
for which we write the integral over TLS frequencies with the unitless $x=\hbar\omega_\textrm{TLS}/(k_B T)$ and group some irrelevant prefactors in $A_d=2^{2-d}\pi^{1-d/2}/\Gamma(d/2)$. This concludes the derivation of the $Q_\textrm{rel}^{-1}\propto T^d$ model that was used in (\ref{eq:relTLS_Q}) of the main text. In the case of a weak energy dependence of the TLS density of states, $\rho(E)\propto E^\mu$, the temperature scaling would be modified to $Q_\textrm{rel}^{-1}\propto T^{d+\mu}$. Importantly, as temperature is increased, the thermal phonon wavelength $\lambda_\textrm{th}=h\overline{v}/(k_B T)$ decreases and may become comparable to the system's dimensions, so that a reduced phonon-bath dimensionality may no longer be guaranteed, as was discussed in Ref.~ \cite{Wollack2021}. For $d=3$, $\Gamma(3/2)=\sqrt{\pi}/2$, $A_3=1/\pi$ and taking $x_\textrm{max}\to\infty$, $\int_{0}^\infty x^3 \textrm{csch}(x)\textrm{d}x=\pi^4/8$, we recover the result derived by Phillips \cite{Phillips1987}, up to a factor of $3$ that accounts for orientational average $\langle M^2\rangle_{xyz}=\overline{M}^2/3$:
\begin{equation}
    Q_\textrm{rel,d=3}^{-1}=\frac{\pi^2}{8}\frac{F\delta_0}{\rho\omega_r\hbar^4}\frac{\overline{D}^2}{\overline{v}^5}(k_B T)^3.
\end{equation}

Although we do not consider it in the main text -- for reasons we will shortly explain -- we can compute the resonator frequency shift due to relaxation TLS by taking the real part of the diagonal susceptibility
\begin{align}
    \frac{\delta\omega_r}{\omega_r} &= -PV \int_{0}^{\omega_\textrm{max}}\mathrm{d}\omega_\textrm{TLS}\,\frac{\hbar^2\overline{g_z}^2}{\omega_r k_B T}\Bigl(\frac{\Gamma_1(\omega_\textrm{TLS})}{\omega_r}\Bigr)^2\textrm{sech}^2\frac{\hbar\omega_\textrm{TLS}}{2 k_B T}\nonumber\\
    &=-\frac{A_d^2\overline{D}^2\overline{M}^4 P}{8\hbar^2 \rho^3\overline{v}^{2d+6}\omega_r^2 S_{3-d}^2}\Bigl(\frac{k_B T}{\hbar}\Bigr)^{2d}\int_{0}^{\quad\mathclap{x_\textrm{max}}}\mathrm{d}x\,\textrm{csch}^2(x/2)x^{2d}.
\end{align}
For $d=3$, $\Gamma(3/2)=\sqrt{\pi}/2$, $A_3=1/\pi$ and taking $x_\textrm{max}\to\infty$, $\int_{0}^\infty x^6 \textrm{csch}^2(x/2)\textrm{d}x=64\pi^6/21$, the frequency shift is given by:
\begin{equation}
    \frac{\delta\omega_r}{\omega_r} = -\frac{8\pi^3}{21}\frac{F\delta_0}{(\rho\hbar\omega_r)^2}\frac{\overline{M}^2}{\overline{v}^5}\frac{\overline{D}^2}{\overline{v}^5}\Bigl(\frac{k_B T}{\hbar}\Bigr)^6.
\end{equation}
Because of the large exponents appearing in these expressions, the value of the frequency shift is highly sensitive to temperature and to the speed of sound. Using typical values for quartz $\omega_r=2\pi\times 500~$MHz, $\rho = 2650~$kg/m\textsuperscript{3}, $\overline{v}=4250~$m/s, $\overline{M}=\overline{D}=1~$eV, we show that at mK temperature the frequency shift from TLS relaxation is completely negligible compared to the one from resonant TLS: $\delta\omega_r/\omega_r\sim 10^{-8} F\delta_0$ at $T=100~$mK, $\sim 10^{-14} F\delta_0$ at $T=10~$mK. Only for $T>1.8~$K does $\delta\omega_r/\omega_r\sim F\delta_0$.

\subsection{Derivation of the kinetic master equation}\label{sec:mastereq}

To derive the relaxation term arising from the longitudinal susceptibility in (\ref{chiDiag}), we assumed the kinetic master equation (\ref{rho}). This simple form describing the relaxation of the TLS density operator towards an equilibrium value fixed by the phonon bath temperature may have appeared \textit{ad hoc}. For full rigor, we demonstrate in this section that it can be derived from the Lindblad master equation. This exercise will also allow us to highlight the approximations and assumptions beyond (\ref{rho}) that were used to derive the TLS relaxation damping contribution in section~\ref{sec:acoustSuscep}. 

The Lindblad master equation applied to a single TLS is the well-known quantum optical master equation \cite{Petruccione2002}: 
\begin{align}\label{lindbladeq}
    \frac{\partial}{\partial t}\rho(t) &= -\frac{i}{\hbar}[H, \rho]+\sum_i \Gamma_i\left(L_i\rho L_i^\dag-\frac{1}{2}\Big\{L_i^\dag L_i,\rho\Big\}\right)\\
    &= (N+1)\Gamma\left(\sigma_-\rho(t)\sigma_+-\frac{1}{2}\sigma_+\sigma_-\rho(t)-\frac{1}{2}\rho(t)\sigma_+\sigma_-\right)\nonumber\\
    &+ N\Gamma\left(\sigma_+\rho(t)\sigma_--\frac{1}{2}\sigma_-\sigma_+\rho(t)-\frac{1}{2}\rho(t)\sigma_-\sigma_+\right)\nonumber,
\end{align}
where $H=-(E/2)\sigma_z$ is the TLS Hamiltonian and we consider only the two dissipators $L_- = \sigma_-$ and $L_+=\sigma_+$. They come with the following damping rates: $\Gamma_-=(N+1)\Gamma$ and $\Gamma_+=N\Gamma$, describing TLS spontaneous emission processes (rate $\Gamma$) as well as thermally induced emission and absorption processes (rates $N\Gamma$), where $N=N(E)=1/(e^{E/k_B T}-1)$ denotes the Planck distribution of phonons at the TLS frequency. In the second line we discard the commutator term describing the Hamiltonian evolution as we are only interested in the dissipative part of the evolution to derive the relaxation term in (\ref{rho}).

With the usual Pauli algebra, (\ref{lindbladeq}) can be recast into a set of linear (Bloch) equations for the populations of the TLS ground and excited state levels (respectively $\rho_{11}(t)=p_g(t)$ and $\rho_{22}(t)=p_e(t)$) and the coherences ($\rho_{12}(t)=\rho_{21}^*(t)$):
\begin{align}
    \frac{\partial}{\partial t}\rho_{11}(t) &= \Gamma_-\rho_{22}(t)-\Gamma_+\rho_{11}(t)\\
    \frac{\partial}{\partial t}\rho_{22}(t) &= -\Gamma_-\rho_{22}(t)+\Gamma_+\rho_{11}(t)\nonumber\\
    \frac{\partial}{\partial t}\rho_{12}(t) &= -\Gamma_\phi\rho_{12}(t)\nonumber,
\end{align}
where we introduce the coherence damping rate $\Gamma_\phi=(N+1/2)\Gamma$. The stationary solution for the populations is obtained by setting $\partial/\partial t=0$: $\rho_{22,\infty} = (\Gamma_+/\Gamma_-)\rho_{11,\infty}=N/(2N+1)$ and $\rho_{11,\infty}=(N+1)/(2N+1)$. These stationary solutions coincide with the results expected from thermal equilibrium; plugging the expression for the phonon number $N(E)$, one recovers $\rho_{22,\infty}=1/(1+e^{E/k_B T})=\rho_{e,th}=f_2$.
We thus rewrite the Bloch equations as:
\begin{align}
    \frac{\partial}{\partial t}\rho_{11}(t) &= -\Gamma_+\left(\rho_{11}(t)-\frac{\Gamma_-}{\Gamma_+}\rho_{22}(t)\right)\\
    &\approx-\Gamma_+\left(\rho_{11}(t)-\rho_{11,\infty}(t)\right)\nonumber\\
    \frac{\partial}{\partial t}\rho_{22}(t) &= -\Gamma_-\left(\rho_{22}(t)-\frac{\Gamma_+}{\Gamma_-}\rho_{11}(t)\right)\nonumber\\
    &\approx-\Gamma_-\left(\rho_{22}(t)-\rho_{22,\infty}(t)\right)\nonumber\\
    \frac{\partial}{\partial t}\rho_{12}(t) &= - \Gamma_\phi\rho_{12}(t)\nonumber,
\end{align}
where we approximate the second term in the right-hand side of the equations for the populations by their instantaneous quasi-equilibrium values. Finally, we express the set of equations more compactly as \footnote{The product in the right hand side of (\ref{kineticEq}) is not the standard matrix product but an element-wise multiplication (Hadamard product).}
\begin{equation}\label{kineticEq}
    \frac{\partial}{\partial t}\hat{\rho}(t) = -\hat{\Gamma}\left(\hat{\rho}(t)-\hat{\rho}_{th}(t)\right), 
\end{equation}
for which we introduce the damping matrix and the thermal equilibrium density matrix
\begin{equation}
    \hat{\Gamma}=\begin{pmatrix}
        \Gamma_+ & \Gamma_\phi\\
        \Gamma_\phi & \Gamma_-
    \end{pmatrix},\quad
    \hat{\rho}_{th}=\begin{pmatrix}
        \rho_{g,th} & 0\\
        0 & \rho_{e,th}
    \end{pmatrix}.
\end{equation}
The right-hand side of (\ref{kineticEq}) is nothing other than the relaxation term in the master equation (\ref{rho}) that was used to derive the general expression for the susceptibility given in (\ref{chi}). 

\section{Dissipative and reactive TLS measurements}\label{sec:apdx_TLSVariance}
\subsection{Variance}
In the main text we consider two methods of measuring the TLS bath which yield the \textit{reactive} and \textit{dissipative} TLS loss tangents \FReac and $F\delta_0^\mathrm{diss}$, respectively. If the TLS distribution is nonuniform in energy spacing or coupling strength, these measurements are not guaranteed to yield the same quantity because the largest contributions to \FDiss arise from near-resonant TLSs while \FReac is affected by far off-resonant TLSs. In other words, these quantities sample different populations of the TLS bath, therefore inhomogeneities in the TLS distribution will cause a disagreement in the reactive and dissipative loss tangents. Here, we derive the relative variance $\mathrm{Var}[F\delta_0^\mathrm{diss}]/\mathrm{Var}[F\delta_0^\mathrm{reac}]$ of these measurements given a bath of TLSs with frequencies which are drawn from a uniform distribution $\omega_\mathrm{TLS}^i\in[ 0, \omega_\mathrm{max}]$. We begin by rewriting the loss tangents in terms of the zero-temperature susceptibility $\chi_0=\chi(T=0)$,
\begin{equation}
    \begin{aligned}
        F\delta_0^\mathrm{diss} &= \frac{2\xi_\mathrm{vac}^2}{\hbar\omega_r} \Im\{\chi_0\} \\
        F\delta_0^\mathrm{reac} &= \frac{\pi\xi_\mathrm{vac}^2}{\hbar\omega_r}\frac{1}{\mathrm{ln}(\omega_\mathrm{max}/\omega_r)}\Re\{\chi_0\},
    \end{aligned}
\end{equation}
which we obtain by combining (\ref{eq:apdx_acousticChi_fshiftTLS}) and (\ref{eq:apdx_acousticChi_kappaTLS}) with (\ref{eq:apdx_acousticChi_domegar}) and (\ref{eq:apdx_acousticChi_QTLS}). We recast the non-diagonal susceptibility from (\ref{eq:apdx_acousticChi_chiNDSum}) as a sum over discrete TLSs and taking the limit of zero temperature, $\chi_0 =\sum_i \chi_0^i$, where the susceptibility contribution $\chi_o^i$ of the $i$-th TLS is
\begin{equation}
    \chi_0^i = \frac{\hbar g_x^2}{\xi_\textrm{vac}^2} \left(\frac{1}{\Delta_i-i\Gamma_{2}}+\frac{1}{\Delta_i+2\omega+i\Gamma_{2}}\right),
\end{equation}   
and $\Delta_i=\omega_\textrm{TLS}^i-\omega_r$ is the TLS-resonator detuning. Applying Bienaymé's identity,
\begin{equation}
    \begin{aligned}
        \mathrm{Var}[\Im\{\chi_0\}] &= \sum_i \mathrm{Var[\Im\{\chi_0^i\}]} \\ 
        \mathrm{Var}[\Re\{\chi_0\}] &= \sum_i \mathrm{Var[\Re\{\chi_0^i\}]}, \\ 
    \end{aligned}
\end{equation}
for which
\begin{multline}
    \mathrm{Var[\Im\{\chi_0^i\}]} = \frac{1}{\omega_\mathrm{max}}\int_{-\omega}^{\omega_\mathrm{max}-\omega}  \Im\{\chi_0^i\}^2 d\Delta_i \\ 
    -\frac{1}{\omega_\mathrm{max}^2}\left(\int_{-\omega}^{\omega_\mathrm{max}-\omega} \Im\{\chi_0^i\}  d\Delta_i\right)^2 \\
    \mathrm{Var[\Re\{\chi_0^i\}]} = \frac{1}{\omega_\mathrm{max}}\int_{-\omega}^{\omega_\mathrm{max}-\omega}  \Re\{\chi_0^i\}^2 d\Delta_i \\ 
    -\frac{1}{\omega_\mathrm{max}^2}\left(\int_{-\omega}^{\omega_\mathrm{max}-\omega} \Re\{\chi_0^i\}  d\Delta_i\right)^2. \\
\end{multline}
Expanding $\Gamma_2$ to first order in the limit that $\omega_\mathrm{max}\rightarrow\infty$, we find
\begin{multline}
    \frac{\mathrm{Var[\Im\{\chi_0\}]}}{\mathrm{Var[\Re\{\chi_0\}]}} = 1 + \frac{8-2\pi^2+8\mathrm{ln^2\bigl(\omega_r/\omega_\mathrm{max}\bigr)}}{\pi \omega_\mathrm{max}}\Gamma_2 \\
    +\mathcal{O}((\Gamma_2)^2).
\end{multline}
Now solving for the relative variance of the TLS loss tangents in the weak TLS damping limit $\Gamma_2\ll\omega_\mathrm{max}$, we see
\begin{equation}
    \frac{\mathrm{Var}[F\delta_0^\mathrm{diss}]}{\mathrm{Var}[F\delta_0^\mathrm{reac}]}\approx\frac{4}{\pi^2}\mathrm{ln}\left(\frac{\omega_\mathrm{max}}{\omega_r}\right)^2.
\end{equation}
Assuming that $\omega_\mathrm{max}\gg\omega_r$, this expression suggests that \FDiss will demonstrate larger variance than \FReac for random distributions of the TLS energies, and \FReac will yield a better measurement of the `true' average TLS ensemble properties. Conversely, if the goal of the measurement is to probe TLSs with energies near $\hbar\omega_r$, then \FDiss is more sensitive to local TLS properties. Although it is difficult to make empirical statements about $\omega_\mathrm{max}$, this analysis suggests that the discrepancy between \FReac and \FDiss may be particularly large for low-frequency resonators.

\subsection{Dissipative measurements}
\begin{figure}
\centering
\includegraphics[width=\columnwidth]{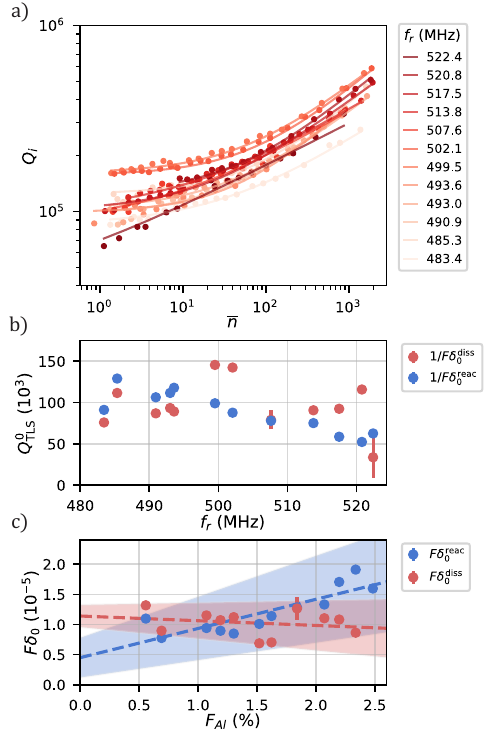}
\caption{Dissipative measurements of the TLS bath. (a) Internal quality factors as a function of driven intracavity phonon occupancy $\overline{n}$ with fits to the resonant TLS loss model. (b) The predicted zero-temperature, zero-drive power internal quality factor $Q_\mathrm{TLS}^0=1/F\delta_0$ from the fitted TLS loss tangents of both reactive and dissipative measurements. (c) The reactive and dissipative loss tangents plotted against the simulated aluminum participation $F_{Al}$.}
\label{fig:apdx_dissVsReac}    
\end{figure}

In the main text, we measured \FReac for 12 of the PCR resonances. Here, we show measurements of \FDiss for the same 12 resonances which were taken immediately prior to the reactive measurements. In this experiment, the refrigerator base plate temperature is maintained at $T=8$ mK while the drive power $P_s$ is swept. Beginning with drive power which corresponds to $\overline{n}=2,000$ for each resonance, we decrement $P_s$ by one dB until the circulating field reaches single-phonon occupancies. At such small powers the PCRs are typically undercoupled by a factor 100-300 which results in a low SNR and necessitates long averaging times. To optimize the measurement process and reduce fit uncertainty, we implement homophasal point distribution \cite{Baity2023, Ganjam2023} and dynamically adjust the number of trace averages to stabilize the SNR (Appendix \ref{sec:apdx_fittingRoutines}). We extract $Q_i$ from each of these traces and fit the results to (\ref{eq:resTLS_Q}); the fits are shown in Fig. \ref{fig:apdx_dissVsReac}(a). In contrast to the measurements of $\Delta f/f_0$ from the main text which were well-stratified by resonance frequency, the traces of $Q_i(\overline{n})$ are largely overlapping and do not show an obvious stratification. We quantify this by plotting in Fig. \ref{fig:apdx_dissVsReac}(b) the predicted zero-temperature zero-power quality factors $Q_\mathrm{TLS}^0=1/F\delta_0$ for both \FDiss and \FReac as shown in the main text. In Fig. \ref{fig:apdx_dissVsReac}(c) we plot $F\delta_0$ against the simulated mechanical aluminum participation $F_{Al}$ for each resonance with a linear fit to the loss tangents. We exclude the data for the $522.4$ MHz resonance which is poorly constrained by the model we fit to. 

Interestingly, the fit to \FDiss is not increasing with $F_{Al}$ as is the case for $F\delta_0^\mathrm{reac}$. We consider two possible reasons for this: the relative variance of the measurement techniques and their respective sensitivities to local TLS density fluctuations. Foremost, as derived in the previous section, we expect a larger variance to arise from the dissipative measurement of the TLS loss tangent given random realizations of the TLS energies. In other words, the disagreement between \FReac and \FDiss may be a predictable statistical consequence of the fact that \FReac is affected by far off-resonant TLSs whereas \FDiss is not. It is also possible, however, that the TLS energy density is inherently nonuniform and that although the ensemble global density is increasing with the aluminum participation, the local TLS density around the resonator is uncorrelated. Given the larger 
anticipated variance associated with \FDiss we assume the former argument, but cannot rule out the latter with these data. Future studies could explore this further by heating and cooling a device many times to `reset' the TLS ensemble while monitoring the means and variances of the dissipative and reactive loss tangents.

\bibliography{biblio}

\end{document}